\RequirePackage{ifpdf}
\ifpdf 
\documentclass[pdftex]{sigma}
\else
\documentclass{sigma}
\fi

\usepackage{euscript}

\newcommand*{\hm}[1]{#1\nobreak\discretionary{}%
            {\hbox{$\mathsurround=0pt #1$}}{}}

\numberwithin{equation}{section}

\newcommand{\End}{\mathop{\mathrm{End}}\nolimits}
\newcommand{\tr}{\mathop{\mathrm{tr}}\nolimits}
\newcommand{\id}{\mathop{\mathrm{id}}\nolimits}
\newcommand{\wt}{\widetilde}
\newcommand{\wh}{\widehat}
\newcommand{\slt}{\mathfrak{sl}_2}
\newcommand{\gln}{\mathfrak{gl}_n}
\newcommand{\A}{{\mathcal A}}

\renewcommand{\L}{{\cal L}}
\renewcommand{\Im}{\mathop{\mathrm{Im}}\nolimits}

\renewcommand{\le}{\leqslant}
\renewcommand{\ge}{\geqslant}

\begin{document}

\allowdisplaybreaks

\renewcommand{\thefootnote}{$\star$}

\renewcommand{\PaperNumber}{110}

\FirstPageHeading

\ShortArticleName{Manin Matrices, Quantum Elliptic Commutative Families and Elliptic Gaudin Model}

\ArticleName{Manin Matrices, Quantum Elliptic Commutative\\ Families and Characteristic Polynomial\\ of Elliptic Gaudin Model\footnote{This paper is a contribution to the Proceedings of the Workshop ``Elliptic Integrable Systems, Isomonodromy Problems, and Hypergeometric Functions'' (July 21--25, 2008, MPIM, Bonn, Germany). The full collection
is available at
\href{http://www.emis.de/journals/SIGMA/Elliptic-Integrable-Systems.html}{http://www.emis.de/journals/SIGMA/Elliptic-Integrable-Systems.html}}}

\Author{Vladimir RUBTSOV~$^{\dag\S}$, Alexey SILANTYEV~$^\ddag$ and Dmitri TALALAEV~$^\S$}

\AuthorNameForHeading{V.~Rubtsov, A.~Silantyev and D.~Talalaev}

\Address{$^\dag$~LAREMA, Universit\'e d'Angers, 2 Boulevard Lavoisier, 49045 Angers, France}
\EmailD{\href{mailto:volodya@univ-angers.fr}{volodya@univ-angers.fr}}
\URLaddressD{\url{http://www.math.univ-angers.fr/~volodya/}}

\Address{$^\ddag$~Department of Mathematics, University Gardens, University of Glasgow, G12 8QW, UK}
\EmailD{\href{mailto:asilantyev@maths.gla.ac.uk}{asilantyev@maths.gla.ac.uk}}

\Address{$^\S$~ITEP, B. Cheremushkinskaja 25, 117218 Moscow, Russia}
\EmailD{\href{mailto:talalaev@itep.ru}{talalaev@itep.ru}}

\ArticleDates{Received March 30, 2009, in f\/inal form December 12, 2009; Published online December 24, 2009}

\Abstract{In this paper we construct the quantum spectral curve for the quantum {\it dynamical} elliptic $\mathfrak{gl}_n$ Gaudin model. We realize it considering a commutative family corresponding to the Felder's elliptic quantum group $E_{\tau,\hbar}(\mathfrak{gl}_n)$ and taking the appropriate limit. The approach of Manin matrices here suits well to the problem of constructing the generation function of commuting elements which plays an important role in SoV and Langlands concept.}


\Classification{37K15}

\renewcommand{\thefootnote}{\arabic{footnote}}
\setcounter{footnote}{0}

\section{Introduction}

The Gaudin model plays intriguingly important role in the modern mathematical physics as like as in purely mathematical subjects as the Geometric Langlands correspondence. In fact, it is shown that the separation of variables of the quantum rational Gaudin model is equivalent to the categorical part of the geometric Langlands correspondence over the rational curve with punctures over $\mathbb{C}$.  Its physical importance lies in the condense matter f\/ield, this system provides an example of the spin interacting magnetic chain. This paper deals with a new useful formalism applicable to the dynamical elliptic case of the Gaudin model. This case provides an interpretation for the Langlands correspondence over an elliptic curve. From the physical point of view this case is responsible for the periodic spin chains.

The main strategy of the paper goes throw the classical ideas of~\cite{KS,KR}.
We start with the formalism of $L$-operators corresponding to the Felder ``elliptic quantum groups'' $E_{\tau,\hbar}(\gln)$~\cite{F1,F2}. Using the $RLL$-relations we construct the Bethe elliptic commutative subalgebra from these $L$-operators. To make the proof of commutativity more apparent we use technique of Manin matrices developed in~\cite{ChF,ChFR}. For some particular cases this commutative subalgebra can be found in~\cite{FV2} in a slightly dif\/ferent form. It is worth to mention the detailed work~\cite{ZhKY} described the centre of $E_{\tau,\hbar}(\mathfrak{sl}_n)$. We refer also to the rational version in~\cite{NO,Mol}, to the trigonometric dynamical case in~\cite{ABB,T1}.
Then we degenerate these families to obtain a commutative family for the elliptic $\gln$ Gaudin model\footnote{Sometimes we omit the word ``dynamical'' for briefness.}. To do it we consider a quantum characteristic polynomial, which has a suitable form for the degeneration. This approach extends the method applied in the rational case in~\cite{T}.

The article is organized as follows: in Subsection~\ref{sec11} we consider the dynamical elliptic $L$-operators def\/ined by the $RLL$ relation with the Felder elliptic $R$-matrix. Then in Subsection~\ref{sec12} we def\/ine Manin matrices following~\cite{ChF,ChFR} and demonstrate that the $L$-operator multiplied by some shift operator is a Manin matrix. In Subsection~\ref{sec13} we consider a fusion procedure of the $L$-operators and construct a commutative family as traces of the ``fused'' $L$-operators. In Subsection~\ref{sec14} we show that a characteristic polynomial of the considered Manin matrix generates the obtained commutative family. Subsection~\ref{sec15} is devoted to  commutative families corresponding to the traces of powers of Manin matrices related to the families constructed in~\ref{sec13} via Newton identities. In Subsection~\ref{sec16} we consider brief\/ly a trigonometric degeneration of the characte\-ris\-tic polynomial and of the commutative families. We discuss their connection with the Hopf algebra $U_q(\wh{\gln})$.

In Subsection~\ref{sec21} we obtain a characteristic polynomial which generates a commutative family by passing to the limit $\hbar\to0$. In Subsection~\ref{sec22} we consider $L$-operators that gives commutative families for the elliptic $\gln$ Gaudin model. Subsection~\ref{sec23} is devoted to a twist relating these $L$-operators with standard $L$-operators for the elliptic $\gln$ Gaudin model. We consider $\slt$ case as an example in Subsection~\ref{sec24}. In Subsection~\ref{sec25} we present the result of application of the Newton identities to the elliptic $\gln$ Gaudin model.

\section{Elliptic quantum groups and commutative families}
\label{QA}

Here we use the $\gln$ dynamical $RLL$ relation to construct the commutative families for dynamical $L$-operators. We consider an arbitrary $L$-operator that def\/ines representation of the ``elliptic quantum group'' $E_{\tau,\hbar}(\gln)$ introduced in~\cite{F1,F2}. The commutativity in the dynamical case is understood modulo Cartan elements. Restricting these ``commutative'' families to the space annihilating by Cartan elements one obtains a corresponding integrable system.

We start with a def\/inition of the odd Riemann theta-function related with an elliptic curve def\/ining the corresponding $R$-matrix. Let $\tau\in\mathbb C$, $\Im\tau>0$ be a module of the elliptic curve~$\mathbb C/\Gamma$, where $\Gamma=\mathbb Z+\tau\mathbb Z$ is a period lattice. The odd theta function $\theta(u)=-\theta(-u)$ is def\/ined as a~holomorphic function on $\mathbb C$ with the properties
\begin{gather*}
 \theta(u+1) =-\theta(u), \qquad \theta(u+\tau) =-e^{-2\pi i u-\pi i\tau}\theta(u), \qquad \theta'(0) =1.
\end{gather*}

We use the matrix (or ``Leningrad'') notations. Let $T=\sum_j t_j\cdot a_{1,j}\otimes\cdots\otimes a_{N,j}$ be a tensor over a ring $\mathfrak R$ (a complex algebra or the f\/ield $\mathbb C$), where $t_j\in\mathfrak R$ and $a_{i,j}$ belong to a space $\End\mathbb C^n$. Then the tensor $T^{(k_1,\ldots,k_N)}$ is the following element of $\mathfrak R\otimes(\End C^n)^{\otimes M}$ for some $M\ge N$:
\begin{gather*}
T^{(k_1,\ldots,k_N)}=\sum_j t_j\cdot1\otimes\cdots\otimes a_{1,j}\otimes\cdots\otimes a_{N,j}\otimes\cdots\otimes1,
\end{gather*}
where each element $a_{i,j}$ is placed to the $k_i$-th tensor factor, the numbers $k_i$ are pairwise dif\/ferent and $1\le k_i\le M$. (The condition $k_1<\cdots<k_N$ is not implied.)

We also use the following notation. Let $F(\lambda)=F(\lambda_1,\ldots,\lambda_n)$ be a function of $n$ parame\-ters~$\lambda_k$ taking values in an algebra ${\mathfrak A}$: that is $F\colon\mathbb C^n\to{\mathfrak A}$. Then we def\/ine
\begin{gather}
 F(\lambda+P)=F(\lambda_1+P_1,\ldots,\lambda_n+P_n)\nonumber\\
 \qquad{} =
 \sum_{i_1,\ldots,i_n=0}^{\infty}\frac{1}{i_1!\cdots i_n!} \frac{\partial^{i_1+\cdots+i_n}F(\lambda_1,\ldots,\lambda_n)}
 {\partial\lambda_1^{i_1}\cdots\partial\lambda_n^{i_n}}P_1^{i_1}\cdots P_n^{i_n}\label{shift}
\end{gather}
for some $P=(P_1,\ldots,P_n)$, $P_k\in{\mathfrak A}$. We omit here the convergence question considering only such situations where this is the case.

\subsection[Elliptic dynamic $RLL$-relations]{Elliptic dynamic $\boldsymbol{RLL}$-relations}
\label{sec11}

First of all we introduce a notion of a dynamical elliptic $L$-operator corresponding to the Felder $R$-matrix. Using a lemma about the products of these $L$-operators and the particular choice of the $L$-operator we prove the commutativity of the family of operators under consideration.

Let $\{e_i\}$ be a standard basis of $\mathbb C^n$ and $\{E_{ij}\}$ be a standard basis of $\End\mathbb C^n$, that is $E_{ij}e_k=\delta^j_ke_i$.
In~\cite{F1,F2} Felder introduce the following element of $\End \mathbb C^n\otimes\End \mathbb C^n$ depen\-ding meromorphically on the {\it spectral parameter} $u$ and $n$ {\it dynamic parameters} $\lambda_1,\ldots,\lambda_n$:
\begin{gather}
 R(u;\lambda)=R(u;\lambda_1,\ldots,\lambda_n)=\frac{\theta(u+\hbar)}{\theta(u)}\sum_{i=1}^n E_{ii}\otimes E_{ii}
 \nonumber\\
\phantom{R(u;\lambda)=}{}
+\sum_{i\ne j}\left(\frac{\theta(\lambda_{ij}+\hbar)}{\theta(\lambda_{ij})} E_{ii}\otimes E_{jj}
 +\frac{\theta(u-\lambda_{ij})\theta(\hbar)}{\theta(u)\theta(-\lambda_{ij})} E_{ij}\otimes E_{ji}\right), \label{R_def}
\end{gather}
where $\lambda_{ij}=\lambda_i-\lambda_j$. It is called  the {\it Felder's dynamical $R$-matrix}. It satisf\/ies the {\it dynamical Yang--Baxter equation} (DYBE)
\begin{gather}
 R^{(12)}(u_1-u_2;\lambda)R^{(13)}(u_1-u_3;\lambda+\hbar E^{(2)})R^{(23)}(u_2-u_3;\lambda) \nonumber \\
 \qquad{} =R^{(23)}(u_2-u_3;\lambda+\hbar E^{(1)})R^{(13)}(u_1-u_3;\lambda)R^{(12)}
 \big(u_1-u_2;\lambda+\hbar E^{(3)}\big) \label{DYBE}
\end{gather}
and the relations
\begin{gather}
 R^{(21)}(-u;\lambda)R^{(12)}(u;\lambda) =\frac{\theta(u+\hbar)\theta(u-\hbar)}{\theta(u)^2}, \label{R21R12} \\
 \big(E^{(1)}_{ii}+E^{(2)}_{ii}\big)R(u;\lambda) =R(u;\lambda)\big(E^{(1)}_{ii}+E^{(2)}_{ii}\big), \label{EER_REE} \\
 \big(\wh D^{(1)}_\lambda+\wh D^{(2)}_\lambda\big)R(u;\lambda) =R(u;\lambda)
 \big(\wh D^{(1)}_\lambda+\wh D^{(2)}_\lambda\big), \label{DR_RD}
\end{gather}
where $\wh D_\lambda=\sum\limits_{k=1}^n E_{kk}\frac{\partial}{\partial\lambda_k}$,  $\wh D^{(i)}_\lambda=\sum\limits_{k=1}^n E^{(i)}_{kk}\frac{\partial}{\partial\lambda_k}$.
We also should comment that we always mean by $\lambda$ the vector $\lambda_1,\ldots,\lambda_n$ and the expression of the type $\lambda+\hbar E^{(s)}$ as the argument of~\eqref{shift} with $P_i=\hbar E^{(s)}_{ii}.$
The relation~\eqref{EER_REE} is obvious. The relation~\eqref{DR_RD} follows from~\eqref{EER_REE}.

Let $\mathfrak R$ be a $\mathbb C[[\hbar]]$-algebra, $L(u;\lambda)$ be an invertible $n\times n$ matrix over $\mathfrak R$ depending on the spectral parameter $u$ and $n$ dynamical parameters $\lambda_1,\ldots,\lambda_n$. Let $h_1, \ldots, h_n$ be a set of some pairwise commuting elements of $\mathfrak R$ . If the matrix $L(u;\lambda)$ satisf\/ies {\it dynamical $RLL$-relation}
\begin{gather}
 R^{(12)}(u-v;\lambda)L^{(1)}(u;\lambda+\hbar E^{(2)})L^{(2)}(v;\lambda)  \notag \\
 \qquad{} =L^{(2)}(v;\lambda+\hbar E^{(1)}) L^{(1)}(u;\lambda)R^{(12)}(u-v;\lambda+\hbar h), \label{DRLL} \\
 (E_{ii}+h_i)L(u;\lambda)=L(u;\lambda) (E_{ii}+h_i), \label{EhL_LEh}
\end{gather}
then it is called a {\it dynamical elliptic $L$-operator} with {\it Cartan elements} $h_k$. The argument $\lambda+\hbar h$ is always meant in the sense of \eqref{shift} with $P_i=\hbar h_i$.

Let us introduce an equivalent but more symmetric form of $RLL$ relations. For each $L$-operator we introduce the following operator (similar to an operator introduced in~\cite{ABB}):
\begin{equation} \label{L_D}
L_D(u)=e^{-\hbar \wh D_{\lambda}}L(u;\lambda).
\end{equation}
The equation \eqref{DRLL} can be rewritten in terms of this operator:
\begin{gather}
R^{(12)}(u-v;\lambda)L_D^{(1)}(u)L_D^{(2)}(v)
 =L_D^{(2)}(v)L_D^{(1)}(u)R^{(12)}(u-v;\lambda+\hbar h). \label{RLLSym}
\end{gather}

\begin{lemma} \label{lem_LL}
 If $L_1(u;\lambda)\in\End(\mathbb C^n)\otimes\mathfrak R_1$ and $L_2(u;\lambda)\in\End(\mathbb C^n)\otimes\mathfrak R_2$ are two dynamical elliptic $L$-operators subject to the two sets
of Cartan elements: $h^1=(h_1^1,\ldots,h_n^1)$ and $h^2=(h_1^2,\ldots,h_n^2)$
 then their matrix product $L_2(u;\lambda)L_1(u;\lambda+\hbar h^2)\in\End(\mathbb C^n)\otimes\mathfrak R_1\otimes\mathfrak R_2$ is also a dynamical elliptic $L$-operator with Cartan elements $h=h^1+h^2=(h_1^1+h_1^2,\ldots,h_n^1+h_n^2).$
Thus, if $L_1(u;\lambda),\ldots,$ $L_m(u;\lambda)$ are dynamical elliptic $L$-operators with Cartan elements $h^1, \ldots, h^m$ then the matrix
\begin{gather*}
 \mathop{\overleftarrow\prod}
  \limits_{m\ge j\ge1}L_j\left(u;\lambda+\hbar\sum_{l=j+1}^{m} h^l\right)
 \end{gather*}
is a dynamical elliptic $L$-operator with Cartan elements $h=\sum\limits_{i=1}^m h^i$.
\end{lemma}

\begin{remark}
The arrow in the product means the order of the factors with respect to the growing index value: the expression $\mathop{\overleftarrow\prod}\limits_{3\ge i\ge1} A_i$ means $A_3 A_2 A_1.$
\end{remark}

The basic example of the dynamical elliptic $L$-operator is the dynamical Felder $R$-matrix: $L(u)=R(u-v;\lambda)$. In this case the second space $\End(\mathbb C^n)$ plays the role of the algebra $\mathfrak R$. Here $v$ is a f\/ixed complex number and the Cartan elements are $h_k=E^{(2)}_{kk}$. Lemma~\ref{lem_LL} allows to generalize this example: let $v_1, \ldots, v_m$ be f\/ixed numbers, then the matrix
\begin{gather*}
 \mathbb R^{(0)}(u;\{v_j\};\lambda)=\mathop{\overleftarrow\prod}\limits_{m\ge j\ge1}R^{(0j)}\left(u-v_j;\lambda+\hbar\sum_{l=j+1}^{m} E^{(l)}\right)
\end{gather*}
is a dynamical elliptic $L$-operator with Cartan elements $h_k=\sum\limits_{l=1}^m E^{(l)}_{kk}$.

The relation~\eqref{R21R12} gives another important example of the dynamical elliptic $L$-operator -- the matrix $L(u)=R^{(21)}(v-u)^{-1}$ with the second space $\End\mathbb C^n$ considering as $\mathfrak R$ (the number $v$ is f\/ixed), or equivalently, the matrix $L(v)=R^{(12)}(u-v)^{-1}$ with the f\/irst space $\End\mathbb C^n$ considering as $\mathfrak R$. Let us introduce the notation
\begin{gather*}
 \mathbb R^{(0)}(\{u_i\};v;\lambda)=\mathop{\overrightarrow\prod}\limits_{1\le i\le m}R^{(i0)}\left(u_i-v;\lambda+\hbar\sum_{l=i+1}^{m} E^{(l)}\right),
\end{gather*}
where $u_1, \ldots, u_m$ are f\/ixed numbers. Then the matrix $\mathbb R^{(0)}(\{u_i\};v;\lambda)^{-1}$ is a dynamical elliptic $L$-operator with the spectral parameter $v$ and the Cartan elements $h_k=\sum\limits_{l=1}^m E^{(l)}_{kk}$.

  A more general class of the dynamical elliptic $L$-operators associated with {\it small elliptic quantum group} $e_{\tau,\hbar}(\gln)$ was constructed in the work~\cite{TV}. This is an $\mathbb C[[\hbar]]((\lambda_1,\ldots,\lambda_n))$-algebra generated by the elements $\tilde t_{ij}$ and $h_k$ satisfying the commutation relations
\begin{gather*}
 \tilde t_{ij}h_k=(h_k-\delta_{ik}+\delta_{jk})\tilde t_{ij}, \notag \\
 \hbar^{-1}\big(t_{ij}\lambda_k-(\lambda_k-\hbar\delta_{ik})t_{ij}\big)=0, \notag \\
 \hbar^{-2}\big(t_{ij}t_{ik}-t_{ik}t_{ij}\big)=0, \\
 \hbar^{-2}\left(t_{ik}t_{jk}-\frac{\theta(\lambda^{\{1\}}_{ij}
 +\hbar)}{\theta(\lambda^{\{1\}}_{ij}-\hbar)}t_{jk}t_{ik}\right)=0,\qquad i\ne j, \notag \\
 \hbar^{-2}\left(\frac{\theta(\lambda^{\{2\}}_{jl}+\hbar)}{\theta(\lambda^{\{2\}}_{jl})} t_{ij}t_{kl}-\frac{\theta(\lambda^{\{1\}}_{ik}+\hbar)}{\theta(\lambda^{\{1\}}_{ik})}t_{kl}t_{ij}-\frac{\theta(\lambda^{\{1\}}_{ik}+\lambda^{\{2\}}_{jl})\theta(\hbar)} {\theta(\lambda^{\{1\}}_{ik})\theta(\lambda^{\{2\}}_{jl})}t_{il}t_{kj}\right)=0, \quad i\ne k,\  j\ne l,\! \notag
\end{gather*}
where $t_{ij}=\delta_{ij}+\hbar\tilde t_{ij}$, $\lambda^{\{1\}}_{ij}=\lambda_i-\lambda_j$, $\lambda^{\{2\}}_{ij}=\lambda_i-\lambda_j-\hbar h_i+\hbar h_j$ and the elements $h_1,\ldots,h_k,\lambda_1,\ldots$, $\lambda_k$ commute with each other.
The authors of~\cite{TV} consider the matrix $T(-u)$ with the entries
\begin{gather*} 
 T_{ij}(-u)=\theta(-u+\lambda_{ij}-\hbar h_i)t_{ji},
\end{gather*}
Representing this matrix in the form
\begin{align}\label{TLeD}
 T(-u)=\theta(-u)e^{-\hbar\sum\limits_{k=0}^n(h_k+E_{kk})\partial_{\lambda_k}}L_0(u;\lambda)e^{\hbar\sum\limits_{k=0}^n h_k\partial_{\lambda_k}}
\end{align}
we obtain a dynamical elliptic $L$-operator $L_0(u;\lambda)$ over the algebra $\mathfrak T=e_{\tau,\hbar}(\gln)[[\partial_\lambda]]$ with Cartan elements $h=(h_1,\ldots,h_n)$, where $\mathbb C[[\partial_\lambda]]=\mathbb C[[\partial_{\lambda_1},\ldots,\partial_{\lambda_n}]]$, the elements $\partial_{\lambda_k}=\frac{\partial}{\partial\lambda_k}$ commute with $h_i$ and do not commute with $\tilde t_{ij}$.

\subsection[Manin matrices and $L$-operators]{Manin matrices and $\boldsymbol{L}$-operators}
\label{sec12}

The $RLL$ relations allow to construct the Manin matrices and $q$-Manin matrices investigated in~\cite{ChF, ChFR, ChFRS} starting from the $L$-operators. In particular, the dynamical $RLL$ relation with Felder~$R$ matrix leads to the Manin matrices (here $q=1$). We use the properties of these matrices to prove the commutativity of some function family $\hat t_m(u)$ which will be constructed in the next subsection.
We shall suppose in what follows that all matrix entries belong to some non-commutative ring which satisf\/ies basically the conditions
described precisely in~\cite{ChF,ChFR,ChFRS}.

\begin{definition} \label{MMdef}
 An $n\times n$ matrix $M$ over some (in general, non-commutative) ring is called Manin matrix if the elements $a$, $b$, $c$, $d$ of its any $2\times2$ submatrix, where
\begin{gather*}
 M=\begin{pmatrix}
 \ldots & \ldots & \ldots & \ldots &\ldots \\
 \ldots & a      & \ldots & b & \ldots \\
 \ldots & \ldots & \ldots & \ldots &\ldots \\
 \ldots & c      & \ldots & d & \ldots \\
 \ldots & \ldots & \ldots & \ldots &\ldots
 \end{pmatrix},
\end{gather*}
satisfy the relation $ad-da=cb-bc$ and their elements from the same column commute with each other: $ac=ca$, $bd=db$.
\end{definition}

Let $S_m$ be the symmetric group and $\pi\colon S_n\to\End(\mathbb C^n)^{\otimes m}$ be its standard representation: $\pi(\sigma)(v_1\otimes\cdots\otimes v_m)=v_{\sigma^{-1}(1)}\otimes\cdots\otimes v_{\sigma^{-1}(m)}$. Let us introduce the operator
\begin{gather*}
 A_m=\frac{1}{m!}\sum_{\sigma\in S_m}(-1)^{\sigma}\pi(\sigma),
\end{gather*}
where $(-1)^{\sigma}$ is a sign of the permutation $\sigma$. We also use the notations $A^{[k,m]}\equiv A^{(k+1,\ldots,m)}\equiv(A_{m-k})^{(k+1,\ldots,m)}$. For example,
\begin{gather} \label{A12def}
  A^{(12)}=A_2=\frac12\sum_{i\ne j}\big(E_{ii}\otimes E_{jj}-E_{ij}\otimes E_{ji}\big).
\end{gather}
In terms of the matrix~\eqref{A12def}  Def\/inition~\ref{MMdef} can be reformulated as follows. The matrix $M$ is a~Manin matrix if and only if it satisf\/ies the relation
\begin{gather}
 A^{(12)}M^{(1)}M^{(2)}=A^{(12)}M^{(1)}M^{(2)}A^{(12)}. \label{AMM_AMMA}
\end{gather}

\begin{proposition}
 If $L(u;\lambda)$ is a dynamical $L$-operator then the matrix
\begin{gather} \label{MDLop}
M=e^{-\hbar\wh D_\lambda}L(u;\lambda)e^{\hbar\frac{\partial}{\partial u}}=L_D(u)e^{\hbar\frac{\partial}{\partial u}},
\end{gather}
where $\wh D_\lambda=\sum\limits_{k=1}^n E_{kk}\frac{\partial}{\partial\lambda_k}$, is a Manin matrix.
\end{proposition}

\begin{proof} Substituting $u=-\hbar$ to~\eqref{R_def} we obtain the formula
\begin{gather}
 R(-\hbar;\lambda)=\sum_{i\ne j}\frac{\theta(\lambda_{ij}+\hbar)}{\theta(\lambda_{ij})} \big(E_{ii}\otimes E_{jj}-E_{ij}\otimes E_{ji}\big), \label{R_mhbar}
\end{gather}
from which we see $R(-\hbar;\lambda)=R(-\hbar;\lambda)A^{(12)}$. Consider the relation~\eqref{RLLSym} at $u-v=-\hbar$. The multiplication of this relation by $A^{(12)}$ from the right does not change its right hand side, hence this does not change its left hand side:
\begin{gather} \label{RLL_RLLA}
 R^{(12)}(-\hbar;\lambda)L_D^{(1)}(u)L_D^{(2)}(u+\hbar)
 =R^{(12)}(-\hbar;\lambda)L_D^{(1)}(u)L_D^{(2)}(u+\hbar)A^{(12)}.
\end{gather}
Multiplying~\eqref{RLL_RLLA} by $e^{2\hbar\frac{\partial}{\partial u}}$ from the right we obtain
\begin{gather}
 R^{(12)}(-\hbar;\lambda)M^{(1)}M^{(2)}=
 R^{(12)}(-\hbar;\lambda)M^{(1)}M^{(2)}A^{(12)}. \label{RMM_RMMA}
\end{gather}
Let us note that the matrix~\eqref{R_mhbar} is related to $A^{(12)}$ by the formula $B(\lambda)R(-\hbar;\lambda)=A^{(12)}$, where
\begin{gather*}
B(\lambda)=\frac12\sum_{i\ne j}\frac{\theta(\lambda_{ij})}{\theta(\lambda_{ij}+\hbar)} E_{ii}\otimes E_{jj}.
\end{gather*}
So, multiplying~\eqref{RMM_RMMA} by $B(\lambda)$ from the left one obtains~\eqref{AMM_AMMA}.
\end{proof}

\begin{lemma}[\cite{ChF,ChFR}] \label{M_inv}
If $M$ is a Manin matrix invertible from the left and from the right then its inverse $M^{-1}$ is also a Manin matrix.
\end{lemma}

In particular, the matrix inverse to~\eqref{MDLop} having the form
\begin{gather} \label{MDLopIn}
 \big(e^{-\hbar\wh D_\lambda}L(u;\lambda)e^{\hbar\frac{\partial}{\partial u}}\big)^{-1}=e^{-\hbar\frac{\partial}{\partial u}}L(u;\lambda)^{-1}e^{\hbar\wh D_\lambda}
\end{gather}
is a  Manin matrix.

\begin{lemma} \label{lem_AMMM_AMMMA}
If $M$ is a Manin matrix then it satisfies the relations
\begin{gather}
 A^{[m,N]}M^{(m+1)}\cdots M^{(N)}=A^{[m,N]}M^{(m+1)}\cdots M^{(N)}A^{[m,N]}, \label{AMMM_AMMMA} \\
 A^{[m,N]}M^{(N)}\cdots M^{(m+1)}=A^{[m,N]}M^{(N)}\cdots M^{(m+1)}A^{[m,N]}, \label{AMMM_AMMMA_Nm}
\end{gather}
where $m<N$.
\end{lemma}

\begin{proof} The idea of the proof of~\eqref{AMMM_AMMMA} is the following. It is suf\/f\/icient to prove that the left hand side does not change if we multiply it by $(-1)^{\sigma_{i,i+1}}\pi(\sigma_{i,i+1})$ from the right, where $\sigma_{i,i+1}$ is an elementary permutation. But the last fact follows from~\eqref{AMM_AMMA}. (See details in~\cite{ChFRS}.)

Multiplying~\eqref{AMMM_AMMMA} by $(-1)^{\sigma_{lst}}\pi(\sigma_{lst})$ from the both sides, where $\sigma_{lst}$ is a
longest permutation, one yields the relation~\eqref{AMMM_AMMMA_Nm}.
\end{proof}

\begin{corollary}
If $L(u;\lambda)$ is a dynamical elliptic $L$-operator then it satisfies
\begin{gather}
 A^{[m,N]}\mathop{\overrightarrow\prod}\limits_{m+1\le i\le N}L^{(i)}\left(u+\hbar(i-m-1);\lambda+\hbar\sum_{l=i+1}^{N}E^{(l)}\right)\nonumber \\
\qquad {} =A^{[m,N]}\mathop{\overrightarrow\prod}\limits_{m+1\le i\le N}L^{(i)}\left(u+\hbar(i-m-1);\lambda+\hbar\sum_{l=i+1}^{N}E^{(l)}\right)A^{[m,N]}, \label{ALLL_ALLLA}
\\
 A^{[m,N]}\mathop{\overleftarrow\prod}\limits_{N\ge j\ge m+1}L^{(j)}\left(v+\hbar(j-m-1);\lambda+\hbar\sum_{l=j+1}^{N}E^{(l)}\right)^{-1} \nonumber \\
\qquad{} =A^{[m,N]}\mathop{\overleftarrow\prod}\limits_{N\ge j\ge m+1}L^{(j)}\left(v+\hbar(j-m-1);\lambda+\hbar\sum_{l=j+1}^{N}E^{(l)}\right)^{-1}A^{[m,N]}. \label{ALLL_ALLLA_inv_Nm}
\end{gather}
\end{corollary}

\begin{proof} The relations~\eqref{ALLL_ALLLA} follows from the formula~\eqref{AMMM_AMMMA} for Manin matrix~\eqref{MDLop}: $M=e^{-\hbar\wh D_\lambda}L(u;\lambda)e^{\hbar\frac{\partial}{\partial u}}$. The relations~\eqref{ALLL_ALLLA_inv_Nm} follows from~\eqref{AMMM_AMMMA_Nm} for Manin matrix~\eqref{MDLopIn}: $M=e^{-\hbar\frac{\partial}{\partial u}}L(u;\lambda)^{-1}e^{\hbar\wh D_\lambda}$.
\end{proof}

Let us consider the following matrix
\begin{gather}
 \mathbb R^{[m,N]}(\{u_i\};\{v_j\};\lambda)=\!\mathop{\overrightarrow\prod}\limits_{1\le i\le m}\mathbb \! R^{(i)}\!\left(u_i;\{v_j\};\lambda+\hbar\sum_{l=i+1}^{m}E^{(l)}\!\right) =\mathop{\overrightarrow\prod}\limits_{1\le i\le m}\mathop{\overleftarrow\prod}\limits_{N\ge j\ge m+1}R^{(ij)},\!\!\! \label{RprRi}
\end{gather}
where
\[
R^{(ij)}=R^{(ij)}\left(u_i-v_j;\lambda+\hbar\sum_{l=i+1}^{m}E^{(l)}+\hbar\sum_{l=j+1}^{N} E^{(l)}\right).
 \]
 Taking into account~\eqref{EER_REE} we obtain the relations $R^{(ij)}R^{(ab)}=R^{(ab)}R^{(ij)}$, where $1\le i<a\le m$, $m+1\le j<b\le N$. Using it we derive the formula
\begin{gather}
 \mathbb R^{[m,N]}(\{u_i\};\{v_j\};\lambda)=\mathop{\overleftarrow\prod}\limits_{N\ge j\ge m+1}\mathbb R^{(j)}\left(\{u_i\};v_j;\lambda+\hbar\sum_{l=j+1}^{N}E^{(l)}\right)\nonumber\\
 \phantom{\mathbb R^{[m,N]}(\{u_i\};\{v_j\};\lambda)}{}
  =\mathop{\overleftarrow\prod}\limits_{N\ge j\ge m+1}\mathop{\overrightarrow\prod}\limits_{1\le i\le m}R^{(ij)}. \label{RprRj}
\end{gather}
Let us set $u_i=u+\hbar(i-1)$, $v_j=v+\hbar(j-m-1)$. Then, substituting $\mathbb R^{(0)}(u;\{v_j\};\lambda)$ and $\mathbb R^{(0)}(\{u_i\};v;\lambda)^{-1}$ as $L$-operators to the formulae~\eqref{ALLL_ALLLA} (with $m=0$, $N=m$) and \eqref{ALLL_ALLLA_inv_Nm} respectively, taking into account the equalities~\eqref{RprRi} and \eqref{RprRj} we obtain
\begin{gather}
A^{[0,m]}\mathbb R^{[m,N]}(u;v;\lambda) =A^{[0,m]}\mathbb R^{[m,N]}(u;v;\lambda)A^{[0,m]},  \label{AR_ARA_m}\\
A^{[m,N]}\mathbb R^{[m,N]}(u;v;\lambda) =A^{[m,N]}\mathbb R^{[m,N]}(u;v;\lambda)A^{[m,N]}, \label{AR_ARA_N}
\end{gather}
where $\mathbb R^{[m,N]}(u;v;\lambda)=\mathbb R^{[m,N]}(\{u_i=u+\hbar(i-1)\};\{v_j=v+\hbar(j-m-1)\};\lambda)$.
Similarly, substituting $\mathbb R^{(0)}(\{u_i\};v;\lambda)^{-1}$ and $\mathbb R^{(0)}(u;\{v_j\};\lambda)$ to the formulae~\eqref{ALLL_ALLLA} and \eqref{ALLL_ALLLA_inv_Nm} (with $m=0$, $N=m$)  respectively, we obtain
\begin{gather}
A^{[0,m]}\mathbb R^{[m,N]}(u;v;\lambda)^{-1} =A^{[0,m]}\mathbb R^{[m,N]}(u;v;\lambda)^{-1}A^{[0,m]}, \label{AR_ARA_inv_m} \\
A^{[m,N]}\mathbb R^{[m,N]}(u;v;\lambda)^{-1} =A^{[m,N]}\mathbb R^{[m,N]}(u;v;\lambda)^{-1}A^{[m,N]}. \label{AR_ARA_inv_N}
\end{gather}

\subsection{Commutative families}
\label{sec13}

The integrals of motion for an integrable system related with an $L$-operator are obtained often as coef\/f\/icients of a decomposition for some operator functions. These functions are constructed as traces of tensor products of $L$-operators multiplied by the ``alternator'' $A^{[0,m]}$. In the dynamical case one should consider the tensor product of the operators~\eqref{L_D}. Now we use the facts proved below to establish the commutativity of these operator functions.

Let us f\/irst f\/ix the dynamical elliptic $L$-operator with Cartan elements $h_k$ and assign the notation $L(u;\lambda)$ to this $L$-operator. Its values at points $u$ belong to the algebra $\End\mathbb C^n\otimes\mathfrak R$.

Introduce the following operators
\begin{gather*}
 \mathbb L^{[m,N]}(\{u_i\};\lambda)=e^{-\hbar\wh D^{(m+1)}_\lambda}L^{(m+1)}(u_{m+1};\lambda)\cdots e^{-\hbar\wh D^{(N)}_\lambda}L^{(N)}(u_N;\lambda),
\end{gather*}
where $m<N$. Using the dynamical $RLL$ relations~\eqref{DRLL} and taking into account~\eqref{DR_RD} we derive the commutation relations for them:
\begin{gather}
 \mathbb R^{[m,N]}\left(\{u_i\};\{v_j\};\lambda-\hbar\sum_{l=1}^N E^{(l)}\right)\mathbb L^{[0,m]}(\{u_i\};\lambda)\mathbb L^{[m,N]}(\{v_j\};\lambda) \nonumber \\
 \qquad{} =\mathbb L^{[m,N]}(\{v_j\};\lambda)\mathbb L^{[0,m]}(\{u_i\};\lambda)\mathbb R^{[m,N]}(\{u_i\};\{v_j\};\lambda+\hbar h). \label{DReLLeLL}
\end{gather}
Setting $u_i=u+\hbar(i-1)$, $v_j=v+\hbar(j-m-1)$ in this relation we obtain the main relation which we need to construct a commutative family:
\begin{gather*}
 \mathbb R^{[m,N]}\left(u;v;\lambda-\hbar\sum_{l=1}^N E^{(l)}\right)\mathbb L^{[0,m]}(u;\lambda)\mathbb L^{[m,N]}(v;\lambda)\nonumber \\
\qquad{} =\mathbb L^{[m,N]}(v;\lambda)\mathbb L^{[0,m]}(u;\lambda)\mathbb R^{[m,N]}(u;v;\lambda+\hbar h), 
\end{gather*}
where we set $\mathbb L^{[a,b]}(u;\lambda)=\mathbb L^{[a,b]}(\{u_i=u+\hbar(i-a-1)\};\lambda)$ for $a<b$.

Let $\mathbb A_n=\mathbb C((\lambda_1,\ldots,\lambda_n))$ be a completed space of functions. The operators $\wh D_\lambda$ acts on the space $\mathbb{A}_n\otimes\mathbb C^n$, so that the operators $\mathbb L^{[a,b]}(u;\lambda)$ act from the space $\mathbb{A}_n\otimes(\mathbb C^n)^{\otimes(b-a)}$ to $\mathbb{A}_n\otimes(\mathbb C^n)^{\otimes(b-a)}\otimes\mathfrak R$: if $u$ is a f\/ixed point then $\mathbb L^{[a,b]}(u;\lambda)\in\End(\mathbb C^n)^{\otimes(b-a)}\otimes\mathfrak A_n$, where $\mathfrak A_n=\mathbb{A}_n[e^{\pm\hbar\partial_{\lambda}}]\otimes\mathfrak R$. Consider the subalgebra $\mathfrak h \subset \mathfrak R\subset\mathfrak A_n$ generated by elements $h_k$ and its normalizer in $\mathfrak A_n$:
\begin{gather*}
 {\mathfrak N}_n={\mathfrak N}_{\mathfrak A_n}(\mathfrak h)=\{x\in \mathfrak A_n \mid \mathfrak h x\subset {\mathfrak A_n}\mathfrak h\}.
\end{gather*}
Let us note that $\mathfrak A_n\mathfrak h$ is a two-side ideal in $\mathfrak N_n$.

\begin{theorem} \label{th_tt_tt}
 Let us define the $\mathfrak A_n$-valued functions
\begin{gather*}
 t_m(u)=\tr\big(A^{[0,m]}\mathbb L^{[0,m]}(u;\lambda)\big),
\end{gather*}
where the trace is implied over all $m$ spaces $\mathbb C^n$.
They commute with Cartan elements $h_k$:
\begin{gather} \label{ht_th}
 h_k t_m(u) =t_m(u)h_k.
\end{gather}
Hence they take values in the subalgebra $\mathfrak N_n$.
And they pair-wise commute modulo the ideal $\mathfrak A_n\mathfrak h\subset\mathfrak N_n$:
\begin{gather} \label{tt_tt}
 t_m(u)t_s(v) =t_s(v)t_m(u)  \qquad {\rm mod} \ \ \mathfrak A_n\mathfrak h.
\end{gather}
\end{theorem}

\begin{proof} The relation~\eqref{ht_th} follows from the formulae
\begin{gather}
 [h_j,\mathbb L^{[0,m]}(u;\lambda)]=-\sum_{l=1}^m [E^{(l)}_{jj},\mathbb L^{[0,m]}(u;\lambda)],\nonumber \\
\sum_{l=1}^m E^{(l)}_{jj}A^{[0,m]}=A^{[0,m]}\sum_{l=1}^m E^{(l)}_{jj} \label{EEEA_AEEE}
\end{gather}
and from the trace periodicity.

Let us stress that the commutativity~\eqref{tt_tt} modulo $\mathfrak A_n\mathfrak h$ is a corollary of the formula
\begin{gather}
 \tr\big(A^{[0,m]}A^{[m,N]}\mathbb L^{[0,m]}(u;\lambda)\mathbb L^{[m,N]}(v;\lambda)\big)\nonumber\\
\qquad{} =\tr\big(A^{[0,m]}A^{[m,N]}\mathbb L^{[m,N]}(v;\lambda)\mathbb L^{[0,m]}(u;\lambda)\mathbb R^{[m,N]}(u;v;\lambda+\hbar h)\mathbb R^{[m,N]}(u;v;\lambda)^{-1}\big),\!\!\label{tt_tt0}
\end{gather}
where $N=m+s$ and the trace is considered over all $N$ spaces $\mathbb C^n$. This formula can be deduced as follows. Consider the left hand side of~\eqref{tt_tt0}. Applying the relation~\eqref{DReLLeLL} we can rewrite it in the form
\begin{gather}
 \tr\Bigg(A^{[0,m]}A^{[m,N]}\mathbb R^{[m,N]}\left(u;v;\lambda-\hbar\sum_{l=1}^N E^{(l)}\right)^{-1}\mathbb L^{[m,N]}(v;\lambda)\nonumber\\
 \qquad{}\times \mathbb L^{[0,m]}(u;\lambda)\mathbb R^{[m,N]}(u;v;\lambda+\hbar h)\Bigg). \label{tt_tt1}
\end{gather}
Note that by virtue of the formula~\eqref{EEEA_AEEE} the relations~\eqref{AR_ARA_inv_m}, \eqref{AR_ARA_inv_N} with the shifted dynamical parameters $\lambda\to\lambda-\hbar\sum\limits_{l=1}^N E^{(l)}$ are still valid. Therefore, we can insert the idempotent $A^{[0,m]}A^{[m,N]}$ after $\mathbb R^{[m,N]}\Big(u;v;\lambda-\hbar\sum\limits_{l=1}^N E^{(l)}\Big)^{-1}$ to the expression~\eqref{tt_tt1}:
\begin{gather*}
 \tr\Bigg(A^{[0,m]}A^{[m,N]}\mathbb R^{[m,N]}\left(u;v;\lambda-\hbar\sum_{l=1}^N E^{(l)}\right)^{-1}A^{[0,m]}A^{[m,N]}\mathbb L^{[m,N]}(v;\lambda)\nonumber \\
\qquad{} \times\mathbb L^{[0,m]}(u;\lambda) \mathbb R^{[m,N]}(u;v;\lambda+\hbar h)\Bigg). 
\end{gather*}
Similarly, taking into account~\eqref{ALLL_ALLLA} we can insert $A^{[0,m]}A^{[m,N]}$ after $\mathbb L^{[0,m]}(u;\lambda)$ and, then, using \eqref{AR_ARA_m}, \eqref{AR_ARA_N} and the trace periodicity we can eliminate $A^{[0,m]}A^{[m,N]}$ before $\mathbb R^{[m,N]}\Big(u;v;\lambda-\hbar\sum\limits_{l=1}^N E^{(l)}\Big)^{-1}$. Thus we obtain
\begin{gather}
 \tr\Bigg(\mathbb R^{[m,N]}\left(u;v;\lambda-\hbar\sum_{l=1}^N E^{(l)}\right)^{-1}A^{[0,m]}A^{[m,N]}\mathbb L^{[m,N]}(v;\lambda)\nonumber \\
\qquad{} \times\mathbb L^{[0,m]}(u;\lambda)A^{[0,m]}A^{[m,N]} \mathbb R^{[m,N]}(u;v;\lambda+\hbar h)\Bigg), \label{tt_tt3}
\end{gather}
where one can delete $A^{[0,m]}A^{[m,N]}$ after $\mathbb L^{[0,m]}(u;\lambda)$. By virtue of~\eqref{EEEA_AEEE} we can present the part after $\mathbb R^{[m,N]}\Big(u;v;\lambda-\hbar\sum\limits_{l=1}^N E^{(l)}\Big)^{-1}$ as $e^{-\hbar\sum\limits_{l=1}^N\wh D^{(l)}_\lambda}Y(\lambda)$, where the matrix $Y(\lambda)$ does not contain the shift operators. Using the formula
\begin{gather*}
 \tr\Bigg(X(\lambda)e^{-\hbar\sum\limits_{l=1}^N\wh D^{(l)}_\lambda}Y(\lambda)\Bigg)= \tr\left(e^{-\hbar\sum\limits_{l=1}^N\wh D^{(l)}_\lambda}Y(\lambda)X\left(\lambda+\hbar\sum_{l=1}^N E^{(l)}\right)\right),
\end{gather*}
where the matrix $X(\lambda)$ does not contain the shift operators and elements of the ring $\mathfrak R$, one can rewrite~\eqref{tt_tt3} as the right hand side of~\eqref{tt_tt0}.
\end{proof}

Thus we obtain the family of functions $t_m(u)$ with values in the algebra $\mathfrak N_n$. Their images by the canonical homomorphism $\mathfrak N_n\to\mathfrak N_n/\mathfrak A_n\mathfrak h$ commute and we obtain the commutative family of functions in the algebra $\mathfrak N_n/\mathfrak A_n\mathfrak h$. Decomposing these functions in some basis functions $\phi_{m,j}(u)$ one yields a commutative family of elements $\hat I_{m,j}$ of this algebra: $t_m(u)=\sum_j\hat I_{m,j}\phi_{k,j}(u)$ mod~$\mathfrak A_n\mathfrak h$. In this way one can construct an integrable system corresponding to a given $L$-operator.

\subsection{Characteristic polynomial}
\label{sec14}

The functions $t_m(u)$ can be gathered to a generating function called the {\it quantum characteristic polynomial}. This polynomial is def\/ined in terms of determinants of Manin matrices. Considering the Gaudin degeneration we shall calculate the degeneration of the characteristic polynomial, although the degeneration of the functions $t_m(u)$ can not be obtained explicitly.

Introduce the following notion of a determinant for a matrix with non-commutative entries. Let $M$ be an arbitrary $n\times n$ matrix, def\/ine its determinant by the formula
\begin{gather*}
 \det M=\sum_{\sigma\in S_n}(-1)^{\sigma}M_{\sigma(1),1}M_{\sigma(2),2}\cdots M_{\sigma(n),n}. 
\end{gather*}
Usually this determinant is called the {\it column determinant}. One can show \cite{ChFRS} that for a Manin matrix $M$ its column determinant coincides with the completely symmetrized determinant and can be represented in the form
\begin{gather}
 \det M=\tr\big(A^{[0,n]}M^{(1)}M^{(2)}\cdots M^{(n)}\big). \label{det_trA}
\end{gather}

\begin{proposition} \label{prop_det_gener}
 Let us consider the Manin matrix $M=e^{-\hbar\wh D_\lambda}L(u;\lambda)e^{\hbar\frac{\partial}{\partial u}}$. The matrix $1-M$ is also a Manin matrix and its determinant generates $t_m(u)$ as follows:
\begin{gather}
 P(u,e^{\hbar\partial_u})=\det\big(1-e^{-\hbar\wh D_\lambda}L(u;\lambda)e^{\hbar\frac{\partial}{\partial u}}\big)=\sum_{m=0}^n(-1)^m t_m(u) e^{m\hbar\frac{\partial}{\partial u}}, \label{det_gener}
\end{gather}
where $t_0(u)=1$. This implies that the determinant~\eqref{det_gener} commutes with $h_k$ as well as that it commutes modulo $\mathfrak A_n{\mathfrak h}$ with itself for different values of the spectral parameter:
\begin{gather*}
 [P(u,e^{\hbar\partial_u}),h_k]  =0, \qquad
[P(u,e^{\hbar\partial_u}),P(v,e^{\hbar\partial_v})] =0\quad {\rm mod} \ \ \mathfrak A_n\mathfrak h.
\end{gather*}
\end{proposition}

\begin{proof} First we note that the functions $t_m(u)$ are related to the principal minors of the mat\-rix~$M$ as follows
\begin{gather}
 \sum_{1\le i_1<\cdots<i_m\le n}\det M^{i_1\ldots i_m}_{i_1\ldots i_m}=\tr_{1,\ldots,m}\big(A^{[0,m]}M^{(1)}\cdots M^{(m)}\big)=t_m(u)e^{m\hbar\frac{\partial}{\partial u}}, \label{det_gener1}
\end{gather}
where $M^{i_1\ldots i_m}_{i_1\ldots i_m}$ is the corresponding submatrix. Due to the formula~\eqref{det_trA} the left hand side of~\eqref{det_gener} decomposes as follows
\begin{gather}
 \det(1-M)=\sum_{m=0}^n(-1)^m\sum_{1\le k_1<\cdots<k_m\le n}\tr_{1,\ldots,n}\big(A^{[0,n]}M^{(k_1)}M^{(k_2)}\cdots M^{(k_m)}\big). \label{det_gener2}
\end{gather}
Using the recursive relation $A^{[0,n]}=\frac1n(A^{[0,n-1]}-(n-1)A^{[0,n-1]}\pi(\sigma_{n-1,n})A^{[0,n-1]})$ one yields
\begin{gather}
 \tr_{1,\ldots,n}\big(A^{[0,n]}M^{(k_1)}M^{(k_2)}\cdots M^{(k_m)}\big)
  =\tr_{1,\ldots,n}\big(A^{[0,n]}M^{(1)}M^{(2)}\cdots M^{(m)}\big)  \label{det_gener3}\\
\phantom{\tr_{1,\ldots,n}\big(A^{[0,n]}M^{(k_1)}M^{(k_2)}\cdots M^{(k_m)}\big)}{}
=\frac{m!(n-m)!}{n!}\tr_{1,\ldots,m}\big(A^{[0,m]}M^{(1)}M^{(2)}\cdots M^{(m)}\big). \nonumber
\end{gather}
The formulae~\eqref{det_gener1}, \eqref{det_gener2}, \eqref{det_gener3} imply~\eqref{det_gener}.
\end{proof}

\subsection{Newton identities and quantum powers}
\label{sec15}

In the theory of classical integrable systems commutative families are usually provided by traces of powers of a classical $L$-operator. Their quantization can be presented as traces of deformed powers of the corresponding quantum $L$-operators. These deformed powers are called {\it quantum powers} of $L$-operator. For rational $L$-operators these quantum powers were described in details in~\cite{ChF}. The main tool to obtain the quantum powers are Newton identities for Manin matrices. They allow to express the quantum powers of $L$-operators through the coef\/f\/icients of the characteristic polynomial.

\begin{lemma}[\cite{ChF,ChFR}]
 Let $M$ be a Manin matrix. Introduce the elements
\begin{gather}
q_m=\tr_{1,\ldots,m}\big(A^{[0,m]}M^{(1)}\cdots M^{(m)}\big)=\sum_{1\le i_1<\cdots<i_m\le n}\det M^{i_1\ldots i_m}_{i_1\ldots i_m}, \label{q_mdef}
\end{gather}
$q_0=1$ and $q_m=0$ for $m>n$. Then for any $m\ge0$ the following identities hold
\begin{gather*}
 mq_m=\sum_{k=0}^{m-1}(-1)^{m+k+1}q_k\tr\big(M^{m-k}\big).
\end{gather*}
They called Newton identities for the Manin matrix $M$.
\end{lemma}

Consider the Manin matrix $M=L_D(u)e^{\hbar\frac{\partial}{\partial u}}=e^{-\hbar\wh D_\lambda}L(u;\lambda)e^{\hbar\frac{\partial}{\partial u}}$. The elements~\eqref{q_mdef} for this matrix have the form $q_m=t_m(u)e^{m\hbar\frac{\partial}{\partial u}}$: see the formula~\eqref{det_gener1}. The powers of $M$ are
\begin{gather*}
 M^k=L_D^{\bf[k]}(u)e^{k\hbar\frac{\partial}{\partial u}}, \qquad \text{where} \qquad L_D^{\bf[k]}(u)=L_D(u)L_D(u+\hbar)\cdots L_D(u+(k-1)\hbar).
\end{gather*}
The matrices $L_D^{\bf[k]}(u)$ are called {\it quantum powers} of the $L$-operator $L_D(u)$. The Newton identities allow recursively express the traces of the quantum powers $\tr\big(L_D^{\bf[k]}(u)\big)$ via the functions $t_m(u)$ and vice versa. Thus we obtain another commutative family of functions generating the same commutative subalgebra in $\mathfrak N_n/\mathfrak A_n\mathfrak h$:
\begin{gather*}
 \big[\tr\big(L_D^{\bf[k]}(u)\big),\tr\big(L_D^{\bf[m]}(v)\big)\big]=0 \qquad {\rm  mod} \ \ \mathfrak A_n\mathfrak h.
\end{gather*}

\subsection{Trigonometric limit}
\label{sec16}

Up to now we have considered $L$-operators satisfying dynamical $RLL$-relations with a certain dynamical elliptic $R$-matrix. But the presented technique is suf\/f\/iciently universal to be applied to many other cases. This approach is directly generalized to the case of arbitrary dynamical $R$-matrix such that $R(-\hbar;\lambda)=B(\lambda)A^{(12)}$ for some invertible matrix $B(\lambda)$. The corresponding $L$-operators give us a family $t_m(u)$ commuting modulo $\mathfrak A_n\mathfrak h$. In particular, it works for the trigonometric and rational degenerations of the Felder $R$-matrix. Using the same scheme for their non-dynamical limits we obtain commuting functions $t_m(u)$ (modulo $0$).

Here we brief\/ly discuss the dynamical and non-dynamical trigonometric limits. The Felder $R$-matrix in the limit $\tau\to i\infty$ takes the form
\begin{gather}
 R_{\rm trig}(u,v;\lambda)=\frac{zq-wq^{-1}}{z-w}\sum_{i=1}^n E_{ii}\otimes E_{ii}\nonumber \\
\phantom{R_{\rm trig}(u,v;\lambda)=}{} +\sum_{i\ne j}\left(\frac{\mu_{ij}q-q^{-1}}{\mu_{ij}-1} E_{ii}\otimes E_{jj}
 +\frac{(q-q^{-1})(z-w\mu_{ij})}{(z-w)(1-\mu_{ij})} E_{ij}\otimes E_{ji}\right),\label{RtrigD}
\end{gather}
where $z=e^{2\pi iu}$, $w=e^{2\pi iv}$, $q=e^{\pi i\hbar}$, $\mu_{kj}=e^{2\pi i\lambda_{kj}}$. An $L$-operator satisfying the relations~\eqref{DRLL} and \eqref{EhL_LEh} with this $R$-matrix~-- a dynamical trigonometric $L$-operator~-- def\/ines the commutative family $t_m(u)$ (modulo corresponding $\mathfrak A_n\mathfrak h$) by means of the formula~\eqref{det_gener}, for instance. One can also consider the commuting traces of quantum powers in the same way as in the Subsection~\ref{sec15}.

Consider the limit $\lambda_k-\lambda_{k+1}\to-i\infty$, that is $\mu_{ij}\to\infty$ for $i<j$ and $\mu_{ij}\to0$ for $i>j$. The limit of the matrix~\eqref{RtrigD} is a non-dynamical trigonometric $R$-matrix
\begin{gather}
 R_{\rm trig}(u-v) =R_{\rm trig}(z,w)=\frac{zq-wq^{-1}}{z-w}\sum_{i=1}^n E_{ii}\otimes E_{ii} + \sum_{i<j}\Bigg(q E_{ii}\otimes E_{jj} +q^{-1} E_{jj}\otimes E_{ii} \notag\\
 \phantom{R_{\rm trig}(u-v) =R_{\rm trig}(z,w)=}{}
  +\frac{(q-q^{-1})w}{(z-w)} E_{ij}\otimes E_{ji}
 +\frac{(q-q^{-1})z}{(z-w)} E_{ji}\otimes E_{ij}\Bigg). \label{Rtrig}
\end{gather}
Let $L(z)$ be an $L$-operator satisfying the usual $RLL$-relations with this matrix:
\begin{gather*}
 R_{\rm trig}(z,w)L^{(1)}(z)L^{(2)}(w)=L^{(2)}(w)L^{(1)}(z)R_{\rm trig}(z,w).
\end{gather*}
The matrix $M=L(z)q^{2z\frac{\partial}{\partial z}}$ is a Manin matrix. The corresponding characteristic polynomial $\det(1-L(u)q^{2z\frac{\partial}{\partial z}})=\sum\limits_{m=0}^n(-1)t_m(z) q^{2mz\frac{\partial}{\partial z}}$  def\/ines a commutative family: $[t_m(z),t_m(w)]=0$. Then, the quantum powers in this case have the form $L^{\bf[k]}(z)=L(z)L(zq^2)\cdots L(zq^{2(k-1)})$, $[\tr\big(L^{\bf[k]}(z)\big),\tr\big(L^{\bf[m]}(w)\big)]=0$.

Let us consider the tensor
\begin{gather}
 \wt R_{\rm trig}(z,w) =F^{(21)}R_{\rm trig}(z,w)F^{-1}=\frac{zq-wq^{-1}}{z-w}\sum_{i=1}^n E_{ii}\otimes E_{ii}
 + \sum_{i<j}\Bigg( E_{ii}\otimes E_{jj}
 \notag\\
 \phantom{\wt R_{\rm trig}(z,w) =}{} + E_{jj}\otimes E_{ii}
 +\frac{(q-q^{-1})w}{(z-w)} E_{ij}\otimes E_{ji}
 +\frac{(q-q^{-1})z}{(z-w)} E_{ji}\otimes E_{ij}\Bigg), \label{Rtildetrig}
\end{gather}
where
\begin{gather*}
 F=q^{\frac12\sum\limits_{i<j}(E_{ii}\otimes E_{jj}-E_{jj}\otimes E_{ii})}=\sum_{i=1}^n E_{ii}\otimes E_{ii}+\sum_{i<j}\Big(q^{\frac12}E_{ii}\otimes E_{jj}+q^{-\frac12} E_{jj}\otimes E_{ii}\Big).
\end{gather*}
This is a standard $R$-matrix for the quasi-triangular Hopf algebra $U_q(\wh{\gln})$. This algebra can be described by two $L$-operators $\wt L^\pm(z)\in\End\mathbb C^n\otimes U_q(\wh{\gln})$ satisfying $RLL$-relation with the $R$-matrix~\eqref{Rtildetrig}. Each $L$-operator $\wt L(z)\in\End\mathbb C^n\otimes\mathfrak R$ satisfying $RLL$-relation with this $R$-matrix def\/ines a homomorphism from the certain Hopf subalgebra $\mathfrak B\subset U_q(\wh{\gln})$ to $\mathfrak R$.\footnote{The algebra $\mathfrak B$ is the algebra described by one of those $L$-operators, for example, by $\wt L^+(z)$. In this case the homomorphism is def\/ined by the formula $\wt L^+_{ij}(z)\mapsto\wt L_{ij}(z)$.} The $R$-matrix~\eqref{Rtildetrig} considered as an $L$-operator def\/ines a representation $\pi_w\colon\mathfrak B\to\End\mathbb C^n$. The subalgebra $\mathfrak B$ contains elements $\hat h_1,\ldots,\hat h_n$ such that $\pi_v(\hat h_k)=E_{kk}$, $\varepsilon(\hat h_k)=0$, $\Delta(\hat h_k)=\hat h_k\otimes1+1\otimes\hat h_k$, where $\varepsilon$ and $\Delta$ are co-unity and co-multiplication of $U_q(\wh{\gln})$. The element
\begin{gather*}
 \mathcal F=q^{\frac12\sum\limits_{i<j}(\hat h_i\otimes \hat h_j-\hat h_j\otimes \hat h_i)}
\end{gather*}
satisf\/ies the cocycle condition (the Drinfeld equation) $\mathcal F^{(12)}(\Delta\otimes\id)(\mathcal F)=\mathcal F^{(23)}(\id\otimes\Delta)(\mathcal F)$, the condition $(\varepsilon\otimes\id)\mathcal F=(\id\otimes\varepsilon)\mathcal F=1$ and, therefore, def\/ines a twist between two co-multiplications of the algebra $U_q(\wh{\gln})$, the standard one $\Delta$ and twisted one $\Delta^{\mathfrak F}$ def\/ined as $\Delta^{\mathfrak F}(x)=\mathcal F\Delta(x)\mathcal F^{-1}$. Let $\phi\colon\mathfrak B\to\mathfrak R$ is the homomorphism def\/ined by a given $L$-operator $\wt L(z)$. Def\/ine the matrix
\begin{gather*}
 G=q^{\frac12\sum\limits_{i<j}(E_{ii}\otimes h_j-E_{jj}\otimes h_i)}=\sum_{i=1}^n q^{\frac12\Big(\sum\limits_{j=i+1}^n h_j-\sum\limits_{j=1}^{i-1}h_j\Big)} E_{ii}.
\end{gather*}
where $h_k=\phi(\hat h_k)$. From the theory of quasi-triangular Hopf algebras it follows that the twisted $L$-operator
$
 L(z)=G\wt L(z)G
$
satisf\/ies $RLL$-relations with the matrix~\eqref{Rtrig}. This means that each $L$-operator $\wt L(z)$ def\/ines the commutative family $t_m(u)$ by means of the characteristic polynomial for the Manin matrix $M=G\wt L(z)Gq^{2z\frac{\partial}{\partial z}}$.

Let us also remark that the commutative families for $\wt L(z)$ can be constructed without twisting. In this case one should use some generalization of the notion of Manin matrices. The matrix $M=\wt L(z)q^{2z\frac{\partial}{\partial z}}$ belongs to such kind of generalization called {\it $q$-Manin matrices} (see~\cite{ChFRS} for details). Nevertheless the approach of $q$-Manin matrices is a little more complicated. Let us also remark that the same situation occurs for the $U_{q,p}(\wh{\gln})$ $L$-operators considered in~\cite{KK}, there exist a dynamical elliptic analogue of the twist $\mathcal F$ relating the $U_{q,p}(\wh{\gln})$ $R$-matrix with the Felder $R$-matrix. So we expect that there exists a generalization of the $q$-Manin matrices corresponding to the dynamical elliptic quantum group~$U_{q,p}(\wh{\gln})$.\footnote{In some formulation $U_{q,p}(\wh{\gln})$ should be interpreted as a Hopf algebroid~\cite{Konno2009}.}

\section{Characteristic polynomial for elliptic Gaudin model}
\label{GAUD}

We consider a degeneration of the dynamical elliptic $RLL$ relations: $\hbar\to0$. In particular, this degeneration describes the {\it dynamical} elliptic $\gln$ Gaudin model. To relate this degenerated $L$-operator with the elliptic Gaudin $L$-operator in the standard formulation we introduce a twist of this $L$-operator. Degenerating the commutative family obtained in the previous section we yield a commutative family for the elliptic Gaudin model. We see that the obtained results generalize the elliptic $\slt$ Gaudin model case investigated in~\cite{EFR,FS}.

\subsection[$\hbar \to 0$ degeneration]{$\boldsymbol{\hbar \to 0}$ degeneration}
\label{sec21}

Here we consider the degeneration of the previous section formulae obtained in the limit $\hbar\to0$: we compare the coef\/f\/icients at the minimal  degree of $\hbar$ that does not lead to a trivial identity. In the degeneration of the dynamical $RLL$-relation, for example, we consider  $\hbar^2$. And the determinant~\eqref{det_gener} is proportional to the $n$-th power of $\hbar$. We avoid using the term the {\it classical degeneration} because we don't want to mismatch the considering degeneration with the classical mechanics degeneration.

Suppose that $L(u;\lambda)$ is a dynamical elliptic $L$-operator of the form
\begin{gather}
 L(u;\lambda)=1+\hbar\L(u;\lambda)+o(\hbar). \label{LqLc}
\end{gather}
where the entries of $\L(u;\lambda)$ belongs to the algebra $\mathfrak R_0=\mathfrak R/\hbar\mathfrak R$. The matrix $\L(u;\lambda)$ is called {\it dynamical classical elliptic $L$-operator}. It satisf\/ies the dynamical classical $rLL$-relations
\begin{gather}
 \big[\L^{(1)}(u;\lambda)-\wh D^{(1)}_\lambda,\L^{(2)}(v;\lambda)-\wh D^{(2)}_\lambda\big]-\sum_{k=1}^n h_k\frac{\partial}{\partial\lambda}r(u-v;\lambda) \nonumber \\
 \qquad{}=\big[\L^{(1)}(u;\lambda)+\L^{(2)}(v;\lambda),r(u-v;\lambda)\big]\label{DrLL}
\end{gather}
with the {\it dynamical classical elliptic} $r$-matrix
\begin{gather}
 r(u;\lambda)=\frac{\theta'(u)}{\theta(u)}\sum_{i=1}^n E_{ii}\otimes E_{ii}
  +\sum_{i\ne j}\left(\frac{\theta'(\lambda_{ij})}{\theta(\lambda_{ij})} E_{ii}\otimes E_{jj}+\frac{\theta(u-\lambda_{ij})}{\theta(u)\theta(-\lambda_{ij})} E_{ij}\otimes E_{ji}\right). \label{cderm}
\end{gather}
The matrix~\eqref{cderm} is related with the Felder $R$-matrix~\eqref{R_def} via the formula
\begin{gather*}
 R(u;\lambda)=1+\hbar r(u;\lambda_\hbar)+o(\hbar).
\end{gather*}
The relation~\eqref{DrLL} follows directly from the relation~\eqref{DRLL}. In turn, the relation~\eqref{EhL_LEh} implies
\begin{gather}
 (E_{ii}+h_i)\L(u;\lambda)=\L(u;\lambda) (E_{ii}+h_i). \label{EhL_LEh_G}
\end{gather}
Let us remark that two dif\/ferent Lie bialgebras described by the relations~\eqref{DrLL} and \eqref{EhL_LEh_G} are considered in great details for the $\slt$-case in~\cite{PRS12}.

\begin{lemma}
 If a Manin matrix has a form $M=1+\hbar{\cal M}+o(\hbar)$ then ${\cal M}$ is also a Manin matrix.
\end{lemma}

\begin{proof} Substituting the decomposition of $M$ to~\eqref{AMM_AMMA} we obtain the equation~\eqref{AMM_AMMA} for ${\cal M}$ at second power of $\hbar$.
\end{proof}

Applying this lemma to the Manin matrix $M=e^{-\hbar\wh D_\lambda}L(u;\lambda)e^{\hbar\frac{\partial}{\partial u}}$ we conclude that the matrix ${\cal M}=\frac{\partial}{\partial u}-\wh D_\lambda+\L(u;\lambda)$ is a Manin matrix. Theorem~\ref{th_tt_tt} and Proposition~\ref{prop_det_gener} give us the following theorem.

\begin{theorem} \label{th_ss_ss}
Let $\A_n={\mathfrak R_0}\otimes\mathbb A_n[\partial_{\lambda}]$ and ${\cal N}_n={\mathfrak N}_{\A_n}(\mathfrak h)=\{x\in \A_n\mid\mathfrak h x\subset\A_n\mathfrak h\}$, where $\mathbb{A}_n=\mathbb C((\lambda_1,\ldots,\lambda_n))$. Define the ${\cal N}_n$-valued functions $s_m(u)$ by the formula
\begin{gather} \label{chpolGaudin}
Q(u,\partial_u)=\det\left(\frac{\partial}{\partial u}-\wh D_\lambda+\L(u;\lambda)\right) =\sum_{m=0}^n s_m(u)\left(\frac{\partial}{\partial u}\right)^{n-m},
\end{gather}
where $s_0(u)=1$. They commute with Cartan elements $h_k$:
\begin{gather} \label{hs_sh}
 h_k s_m(u) =s_m(u)h_k
\end{gather}
and they pair-wise commute  modulo $\A_n\mathfrak h$:
\begin{gather} \label{ss_ss}
 s_m(u)s_l(v) =s_l(v)s_m(u) \qquad {\rm mod} \ \ \A_n\mathfrak h.
\end{gather}
\end{theorem}

The decomposition coef\/f\/icients of the functions $s_1(u), s_2(u), \ldots, s_n(u)$ form a commutative family in the algebra ${\cal N}_n$ at the level $h_k=0$. This means that their images by the canonical homomorphism ${\cal N}_n\to{\cal N}_n/\A_n{\mathfrak h}$ commute with each other.
Below we consider some special class of $L$-operators $\L(u;\lambda)$.

\subsection{Elliptic Gaudin model}
\label{sec22}

Theorem~\ref{th_ss_ss} provides the functions $s_m(u)$ generating the quantum integrals of motion for the {\it elliptic Gaudin model}. This model is def\/ined by a representation of the loop Lie algebra $\gln[z,z^{-1}]$. Let us f\/irst consider a homomorphism $\rho\colon U(\gln[z,z^{-1}])\to\mathfrak R_0$ from $U(\gln[z,z^{-1}])$ to some al\-geb\-ra $\mathfrak R_0$. To this homomorphism we associate the elliptic half-currents
\begin{gather}
 e^+_{ii}(u) =\rho\left(\frac{\theta'(u-z)}{\theta(u-z)}e_{ii}\right)= \sum_{m\ge0}\frac{(-1)^m}{m!}\left(\frac{\theta'(u)}{\theta(u)}\right)^{(m)}\rho(e_{ii}z^{m}), \label{hc_eii} \\
 e^+_{ij}(u;\lambda) =\rho\left(\frac{\theta(u-z+\lambda_{ij})}{\theta(u-z)\theta(\lambda_{ij})}e_{ij}\right)= \sum_{m\ge0}\frac{(-1)^m}{m!}\left(\frac{\theta(u+\lambda_{ij})}{\theta(u)\theta(\lambda_{ij})}\right)^{(m)}\rho(e_{ij}z^{m}), \label{hc_eij}
\end{gather}
where $i\ne j$, $\{e_{ij}\}$ is a basis in $\gln$ for which $e_{ij}\mapsto E_{ij}$ is a faithful representation. The set $\{e_{ii}z^{m}\}$ is a basis in $\gln[z,z^{-1}]$. The matrix $\L(u;\lambda)$ with the elements
\begin{gather} \label{Lcurr}
 \L_{ij}(u;\lambda) =e^+_{ji}(u;\lambda), \qquad \L_{ii}(u;\lambda) =e^+_{ii}(u;\lambda)+\sum_{k\ne i}\frac{\theta'(\lambda_{ik})}{\theta(\lambda_{ik})}h_k,
\end{gather}
where $h_k=\rho(e_{kk})$, satisf\/ies the relations~\eqref{DrLL} and \eqref{EhL_LEh_G}.

Consider an important example of the homomorphism $\rho$. The {\it evaluation homomorphism} $\rho_v\colon U(\gln[z,z^{-1}])\to U(\gln)$ def\/ined as
\begin{gather*}
 \rho_v(e_{ij}z^m)=v^m e_{ij}, 
\end{gather*}
where $i,j=1,\ldots,n$ and $v\in\mathbb C$ is a f\/ixed complex number. Using the standard co-multiplication of the universal enveloping algebra $U(\gln[z,z^{-1}])$ one can construct the more general evaluation homomorphism $\rho_{\{v\}}\colon U(\gln[z,z^{-1}])\to U(\gln)^{\otimes N}$ in $N$ f\/ixed points $v_1,\ldots,v_N\in\mathbb C$:
\begin{gather*} 
 \rho_{\{v\}}(e_{ij}z^m)=\sum_{k=1}^N v_k^m e^{(k)}_{ij}.
\end{gather*}
where $e^{(k)}_{ij}=1^{\otimes(k-1)}\otimes e_{ij}\otimes1^{\otimes(N-k)}$ is the basis element corresponding to the $k$-th tensor component of $U(\gln)$. For the $N$ given representations $\Pi_k\colon U(\gln)\to\End V_k$ one can construct the representation $\Pi_{\{v\}}\colon U(\gln[z,z^{-1}])\to\End V$, where $\Pi_{\{v\}}=(\Pi_1\otimes\cdots\otimes \Pi_N)\circ\rho_{\{v\}}$ and $V=V_1\otimes\cdots\otimes V_N$. In the representation $\Pi_{\{v\}}$ the half currents~\eqref{hc_eii}, \eqref{hc_eij} take the forms
\begin{gather}
 e^+_{ii}(u;\lambda) =e^+_{ii}(u)=\sum_{k=1}^N\frac{\theta'(u-v_k)}{\theta(u-v_k)}\Pi_k(e_{ii})^{(k)}, \label{hc_eii_N} \\
 e^+_{ij}(u;\lambda) =\sum_{k=1}^N \frac{\theta(u-v_k+\lambda_{ij})}{\theta(u-v_k)\theta(\lambda_{ij})}\Pi_k(e_{ij})^{(k)}, \label{hc_eij_N}
\end{gather}
where $\Pi_k(e_{ij})^{(k)}=1^{\otimes(k-1)}\otimes\Pi_k(e_{ij})\otimes1^{\otimes(N-k)}$ (the $k$-th tensor component). The matrix $\L(u;\lambda)$ def\/ined by~\eqref{Lcurr} with the currents~\eqref{hc_eii_N}, \eqref{hc_eij_N} is called {\it the elliptic Gaudin $L$-operator corresponding to the representation} $\Pi_{\{v\}}$.

In the case of homomorphism $\rho=\rho_{\{v\}}$ the corresponding commutative family can be obtained as follows. Let $L_0(u;\lambda)$ be the dynamical elliptic $L$-operator def\/ined by~\eqref{TLeD}. Shifting the argument we again get a dynamical elliptic $L$-operator $L_0(u-v;\lambda)$. This is also a matrix over $\mathfrak T=e_{\tau,\hbar}(\gln)[[\partial_\lambda]]$. Let $\wh{\mathbb A}_n=\mathbb A_n[[\hbar]][[\partial_\lambda]]\subset\mathfrak T$. Let $L_k(u-v_k;\lambda)$ be the matrices over $\mathfrak R=\underbrace{\mathfrak T\otimes_{\wh{\mathbb A}_n}\cdots\otimes_{\wh{\mathbb A}_n}\mathfrak T}_N$ with the entries $\big(L_k(u-v_k;\lambda)\big)_{ij}=1^{\otimes(k-1)}\otimes\big(L_0(u-v_k;\lambda)\big)_{ij}\otimes1^{\otimes(N-k)}$ and $h^k=(h_1^k,\ldots,h_n^k)$ with $h^k_i=1^{\otimes(k-1)}\otimes h_i\otimes1^{\otimes(N-k)}$. Then applying Lemma~\ref{lem_LL} we obtain a dynamical elliptic $L$-operator
$
 L_{[N]}(u;\lambda)=\mathop{\overleftarrow\prod}
  \limits_{N\ge k\ge1}L_k\Big(u-v_k;\lambda+\hbar\sum\limits_{l=k+1}^{N} h^l\Big)
$
over $\mathfrak R$ with the Cartan elements $h=\sum\limits_{k=1}^N h^k$. Representing $L_{[N]}(u;\lambda)$ in the form~\eqref{LqLc} we obtain an $L$-opera\-tor~$\L_{[N]}(u;\lambda)$. This is a matrix over $\mathfrak R_0=\underbrace{\mathfrak T_0\otimes_{\wt{\mathbb A}_n}\cdots\otimes_{\wt{\mathbb A}_n}\mathfrak T_0}_N$, where $\mathfrak T_0=\mathfrak T/\hbar\mathfrak T$ and $\wt{\mathbb A}_n=\mathbb A_n [[\partial_\lambda]]\subset\mathfrak T_0$. Due to Theorem~\ref{th_ss_ss} the functions $s^{[N]}_m(u)$ def\/ined by~\eqref{chpolGaudin} with this $L$-operator $\L_{[N]}(u;\lambda)$ satisfy~\eqref{hs_sh} and~\eqref{ss_ss}. Let $\varphi\colon\mathfrak R_0\to\mathfrak S_0$ be a homomorphism to an algebra $\mathfrak S_0$. Applying it to the equations~\eqref{chpolGaudin}, \eqref{hs_sh} and~\eqref{ss_ss} we conclude that the functions $s^{\varphi}_m(u)=\varphi(s^{[N]}_m(u))$ coincide with the functions def\/ined by~\eqref{chpolGaudin} with the $L$-operator $\L_{\varphi}(u;\lambda)=\varphi\big(\L_{[N]}(u;\lambda)\big)$ and satisfy~\eqref{hs_sh} and~\eqref{ss_ss}.

Consider the homomorphism $\varphi_0\colon\mathfrak T_0\to U(\gln)\otimes\wt{\mathbb A}_n$ def\/ined by the formulae
\begin{gather*}
 \varphi_0(\tilde t_{ii})=\sum_{k\ne i} \frac{\theta'(\lambda_{ik})}{\theta(\lambda_{ik})}e_{kk}-\frac{\partial}{\partial\lambda_i},\qquad
 \varphi_0(\tilde t_{ij})=\frac{1}{\theta(\lambda_{ij})} e_{ij} \quad\text{for} \ \ i\ne j,
\end{gather*}
$\varphi_0(h_k)=e_{kk}$ and $\varphi_0(\partial_{\lambda_k})=\partial_{\lambda_k}$. Let $\varphi_{[N]}\colon\mathfrak R_0\to U(\gln)^{\otimes N}\otimes\wt{\mathbb A}_n$ be a homomorphism def\/ined as $\varphi_{[N]}=\underbrace{\varphi_0\otimes_{\wt{\mathbb A}_n}\cdots\otimes_{\wt{\mathbb A}_n}\varphi_0}_N$. Then the operator $\L_{\varphi_{[N]}}(u;\lambda)=\varphi_{[N]}\big(\L_{[N]}(u;\lambda)\big)$ coincides with the $L$-operator~\eqref{Lcurr} corresponding to the homomorphism $\rho=\rho_{\{v\}}$. The functions $s^{\rho_{\{v\}}}_m(u)$ def\/ined by~\eqref{chpolGaudin} with $\L_{\varphi_{[N]}}(u;\lambda)$ satisfy~\eqref{hs_sh} and \eqref{ss_ss}. Analogously, considering $\varphi=(\Pi_1\otimes\cdots\otimes \Pi_N)\circ\varphi_{[N]}$ we obtain the Gaudin $L$-operator corresponding to the representation $\Pi_{\{v\}}=(\Pi_1\otimes\cdots\otimes \Pi_N)\circ\rho_{\{v\}}$. Thus we derive the following proposition.

\begin{proposition} \label{prop23}
 Let $\Pi_k\colon U(\gln)\to \End V_k$ be representations and $\L(u;\lambda)$ be the elliptic Gaudin $L$-operator corresponding to the homomorphism $\Pi_{\{v\}}=(\Pi_1\otimes\cdots\otimes \Pi_N)\circ\rho_{\{v\}}$. Then $s_m(u)$ defined by the formula~\eqref{chpolGaudin} with this $\L(u;\lambda)$ satisfy~\eqref{hs_sh} and \eqref{ss_ss}. These are integrals of motion for  the dynamical elliptic $\gln$ Gaudin model.
\end{proposition}

So we have shown that the $L$-operator~\eqref{Lcurr} corresponding to the certain representation $\rho\colon U(\gln[z,z^{-1}])\to\mathfrak R_0=\End V$ def\/ines a commutative family $s_m(u)$. We conjecture that this is true for any homomorphism $\rho$.

\begin{conjecture}
  For any homomorphism $\rho$ the corresponding $L$-operator~\eqref{Lcurr} defines a commutative family $s_m(u)$ by the formula~\eqref{chpolGaudin}, that is $s_m(u)$ satisfy~\eqref{hs_sh} and \eqref{ss_ss}.
\end{conjecture}

For the case $n=2$ this conjecture can be proved using a certain dynamical elliptic $L$-operator constructed in~\cite{EF}. In notations of that paper this is the $L$-operator $\bar L(u,\lambda)$ constructed from the $L$-operator $L_\lambda^+(u)$. The entries of $\bar L(u,\lambda)$ belong to the certain algebra $\mathfrak R=U_\hbar\mathfrak g$ (via some embedding) and the degeneration of this $L$-operator gives the $L$-operator~\eqref{Lcurr} for $\rho=\id$.\footnote{The degeneration of this algebra $\mathfrak R=U_\hbar\mathfrak g$ with zero central charge $K=0$ and without co-central charge is $\mathfrak R_0=\mathfrak R/\hbar\mathfrak R=U(\gln[z,z^{-1}])$.} Thus the $L$-operator for $\rho=\id$ def\/ines functions $s^{\id}_m(u)$ satisfying~\eqref{hs_sh} and \eqref{ss_ss}. The $L$-operator for the arbitrary homomorphism $\rho$ def\/ines function $s^{\rho}_m(u)=\rho\big(s_m(u)\big)$, applying $\rho$ to the relations~\eqref{hs_sh} and \eqref{ss_ss} for $s^{\id}_m(u)$ we conclude that $s^{\rho}_m(u)$ satisfy the same relations.

For the general $n$ this conjecture can be proved in the same way. For example, it follows from the Conjecture~5.1 of the paper~\cite{KK}. There the authors present an analogue of the $L$-operator $L^+_\lambda(u)$ for the quantum group $U_{q,p}(\wh{\mathfrak{sl}_n})$. The existence of a twist relating this $L$-operator with an analogue of $\bar L(u,\lambda)$ can be easily proved.

Theorem~\ref{th_ss_ss} with Proposition~\ref{prop23} provide us a method to construct the quantum elliptic Gaudin model for general $n$ in analogy to the rational case treated in~\cite{T}. Let us f\/irstly note that the connection whose determinant $Q(u,\partial_u)$ provides a generating function for quantum Hamilto\-nians is the KZB connection.  Moreover this construction allows to establish important relations of the elliptic Gaudin model with the Langlands correspondence program.
The construction of the characteristic polynomial $Q(u,\partial_u)$ called the {\it Universal $G$-oper} in this case provides us
\begin{itemize}\itemsep=0pt
  \item a universal construction of the commutative family;
	\item a universal description of the eigenvalue problem;
	\item a relation between the solutions of the KZB equation and the wave-functions of the Gaudin model;
	\item a relation with the centre of the af\/f\/ine algebra at the critical level
\end{itemize}
analogously to the results of \cite{CT1,CT2} in the rational case.

Usually the elliptic Gaudin model def\/ined by the $L$-operator
\begin{gather} \label{LopPi}
 \wt\L(u;\lambda)=\sum_{ij=1}^n E_{ij}\otimes e^+_{ji}(u;\lambda)
\end{gather}
with the currents~\eqref{hc_eii_N}, \eqref{hc_eij_N} or, more generally, with the currents~\eqref{hc_eii}, \eqref{hc_eij}. In the next subsection we describe the relation of the $L$-operator~\eqref{LopPi} with the $L$-operator $\L(u;\lambda)$.

\subsection[Twisting of a classical dynamical $L$-operator]{Twisting of a classical dynamical $\boldsymbol{L}$-operator}
\label{sec23}

First we consider the twists of dynamical $r$-matrices for arbitrary Lie algebras. Let $\mathfrak g$ be a (in general, inf\/inite-dimensional) Lie algebra, $\mathfrak g\otimes\mathfrak g$ be a tensor product completed in some way and $\mathfrak h\subset\mathfrak g$ be an $n$-dimensional commutative subalgebra called {\it Cartan subalgebra}. Let us f\/ix a basis $\{\hat h_1,\ldots,\hat h_n\}$ of $\mathfrak h$. An element $\mathfrak r(\lambda)\in\mathfrak g\otimes\mathfrak g$ depending on $n$ dynamical parameters $\lambda_1,\ldots,\lambda_n$ is called a {\it dynamical classical $r$-matrix} if $\mathfrak r^{(12)}(\lambda)+\mathfrak r^{(21)}(\lambda)\in\big(\mathfrak g\otimes\mathfrak g\big)^{\mathfrak g}$ and it satisf\/ies the classical dynamical Yang--Baxter equation
\begin{gather}
 [[\mathfrak r(\lambda),\mathfrak r(\lambda)]]+{\cal D}_\lambda\big(\mathfrak r(\lambda)\big)=0 \label{CDYBE}
\end{gather}
where
\begin{gather}
 [[a(\lambda),b(\lambda)]] =\big[a^{(12)}(\lambda),b^{(13)}(\lambda)\big]+\big[a^{(12)}(\lambda),b^{(23)}(\lambda)\big]+
 \big[a^{(13)}(\lambda),b^{(23)}(\lambda)\big], \label{ab} \\
{\cal D}_\lambda\big(a(\lambda)\big) =-\sum_{i=1}^n \hat h^{(1)}_i\frac{\partial a^{(23)}(\lambda)}{\partial\lambda_i}+\sum_{i=1}^n \hat h^{(2)}_i\frac{\partial a^{(13)}(\lambda)}{\partial\lambda_i}-\sum_{i=1}^n \hat h^{(3)}_i\frac{\partial a^{(12)}(\lambda)}{\partial\lambda_i}. \nonumber 
\end{gather}
The space $\big(\mathfrak g\otimes\mathfrak g\big)^{\mathfrak g}$ consists of the elements $x\in\mathfrak g\otimes\mathfrak g$ such that $[x,y\otimes1+1\otimes y]=0$ for all $y\in\mathfrak g$. The brackets in the right hand side of~\eqref{ab} mean the commutators in $U(\mathfrak g)\otimes U(\mathfrak g)\otimes U(\mathfrak g)$.

\begin{lemma} \label{lem_cdtwist}
Let $\tilde{\mathfrak r}(\lambda)$ and ${\mathfrak f}(\lambda)$ be elements of $\mathfrak g\otimes\mathfrak g$. Suppose that ${\mathfrak f}^{(12)}(\lambda)+{\mathfrak f}^{(21)}(\lambda)\in\big(\mathfrak g\otimes\mathfrak g\big)^{\mathfrak g}$ and
\begin{gather*}
 [[{\mathfrak f}(\lambda),{\mathfrak f}(\lambda)]]+[[\tilde{\mathfrak r}(\lambda),{\mathfrak f}(\lambda)]]+[[{\mathfrak f}(\lambda),\tilde{\mathfrak r}(\lambda)]]+{\cal D}_\lambda\big({\mathfrak f}(\lambda)\big)=0.
\end{gather*}
Then, $\tilde{\mathfrak r}(\lambda)$ is a dynamical classical $r$-matrix if and only if the matrix ${\mathfrak r}(\lambda)=\tilde{\mathfrak r}(\lambda)+{\mathfrak f}(\lambda)$ is a~dynamical classical $r$-matrix. The element ${\mathfrak f}(\lambda)$ is called {\it classical dynamical twist}.
\end{lemma}

Consider the loop algebra $\mathfrak g=\gln[z,z^{-1}]$ with Cartan subalgebra $\mathfrak h$ spanned by the elements $\hat h_k=e_{kk}$. This is the Lie algebra of constant diagonal matrices. Let $\mathfrak r(\lambda)$ be a dynamical classical $r$-matrix and let $\pi_u\colon\gln[z,z^{-1}]\to\End\mathbb C^n$ be the standard evaluation representation def\/ined as $\pi_u\colon e_{ij}z^k\mapsto E_{ij}u^k$. Since $\pi_u$ is faithful the matrix $\mathfrak r$ can be represented by the matrix $r(u,v;\lambda)=(\pi_u\otimes\pi_v){\mathfrak r}(\lambda)\in\End\mathbb C^n\otimes\End\mathbb C^n[[u,u^{-1},v,v^{-1}]]$, where the dependency of~$u$ and~$v$ is understood in the formal way. Each homomorphism $\rho\colon U(\gln[z,z^{-1}])\to\mathfrak R_0$ def\/ines the $L$-operator $\L(u;\lambda)=(\pi_u\otimes\rho)\mathfrak r(\lambda)$ satisfying the dynamical $rLL$-relations~\eqref{DrLL} with the $r$-matrix $r(u,v;\lambda)$.

Let $r(u,v;\lambda)\in\End\mathbb C^n\otimes\End\mathbb C^n[[v]]((u))$ be the $r$-matrix $r(u-v;\lambda)$ (formula~\eqref{cderm}) understood as a formal series with coef\/f\/icients in $\End\mathbb C^n\otimes\End\mathbb C^n$. Relation~\eqref{DYBE} implies that the corresponding element $\mathfrak r(\lambda)$ is a dynamical classical $r$-matrix. The matrix $\L(u;\lambda)=(\pi_u\otimes\rho)\mathfrak r(\lambda)$ coincides with the $L$-operator~\eqref{Lcurr} corresponding to the homomorphism $\rho\colon U(\gln[z,z^{-1}])\to\mathfrak R_0$.

Consider the following twist
\begin{gather}
 {\mathfrak f}(\lambda)=\sum_{i\ne j}\frac{\theta'(\lambda_{ij})}{\theta(\lambda_{ij})} e_{ii}\otimes e_{jj}. \label{cdtwist}
\end{gather}
It satisf\/ies the conditions of the Lemma~\ref{lem_cdtwist}. Indeed, the {\it twisted} $r$-matrix $\tilde{\mathfrak r}(\lambda)=\mathfrak r(\lambda)-{\mathfrak f}(\lambda)$ is represented by the matrix
\begin{gather*}
 \tilde r(u,v;\lambda)=r(u,v\lambda)-(\pi_u\otimes\pi_v){\mathfrak f}(\lambda)
 =\frac{\theta'(u)}{\theta(u)}\sum_{i=1}^n E_{ii}\otimes E_{ii}
  +\sum_{i\ne j}\frac{\theta(u-\lambda_{ij})}{\theta(u)\theta(-\lambda_{ij})} E_{ij}\otimes E_{ji},
\end{gather*}
where $(\pi_u\otimes\pi_v){\mathfrak f}(\lambda)=\sum\limits_{i\ne j}\frac{\theta'(\lambda_{ij})}{\theta(\lambda_{ij})} E_{ii}\otimes E_{jj}$ is the representation of the twist~\eqref{cdtwist}. One can check the equalities $[[{\mathfrak f}(\lambda),{\mathfrak f}(\lambda)]]\hm=[[\tilde{\mathfrak r}(\lambda),{\mathfrak f}(\lambda)]]+[[{\mathfrak f}(\lambda),\tilde{\mathfrak r}(\lambda)]]\hm={\cal D}_\lambda\big({\mathfrak f}(\lambda)\big)\hm=0$ applying homomorphism $(\pi_u\otimes\pi_v)$. So that the matrix $\tilde{\mathfrak r}(\lambda)$ also satisf\/ies~\eqref{CDYBE}.

The {\it twisted} $L$-operator $\wt\L(u;\lambda)=(\pi_u\otimes\rho)\tilde{\mathfrak r}(\lambda)$ satisf\/ies the $rLL$-relations~\eqref{DrLL} with $r$-matrix~$\tilde r(u,v;\lambda)$ and the relation~\eqref{EhL_LEh_G} with $h_k=\rho(\hat h_k)=\rho(e_{kk})$. It coincides with the $L$-operator~\eqref{LopPi} related with the $L$-operator $\L(u;\lambda)$ via the formulae
\begin{gather*} 
 \L_{ij}(u;\lambda) =\wt\L_{ij}(u;\lambda), \qquad  \L_{ii}(u;\lambda) =\wt\L_{ii}(u;\lambda)+\sum_{k\ne i}\frac{\theta'(\lambda_{ik})}{\theta(\lambda_{ik})}h_k,
\end{gather*}
where $j\ne i$. The characteristic polynomial from Theorem~\ref{th_ss_ss} in these terms takes the form
\begin{gather} \label{Q_Ltilde}
Q(u,\partial_u)=\det\left(\frac{\partial}{\partial u}-\wh D_\lambda+\wt\L(u;\lambda)+\sum_{i\ne j}E_{ii}\frac{\theta'(\lambda_{ij})}{\theta(\lambda_{ij})}h_j\right).
\end{gather}

\begin{remark}
Calculating determinant~\eqref{Q_Ltilde} modulo $\A_n\mathfrak h[[u^{-1},u]][\partial_u]$ in the $n=1$ and $n=2$ cases we can omit the sum $\sum\limits_{i\ne j}E_{ii} \frac{\theta'(\lambda_{ij})}{\theta(\lambda_{ij})}h_j$ in the determinant. However for $n\ge3$ one can not simplify it in this way, because even in the case $n=3$ omitting this sum we lose the following term of $Q(u,\partial_u)$ (mod $\A_n\mathfrak h[[u^{-1},u]][\partial_u]$):
\begin{gather*}
 -2\left(\frac{\theta'(\lambda_{12})}{\theta(\lambda_{12})}
 +\frac{\theta'(\lambda_{23})}{\theta(\lambda_{23})}\right)\wt\L_{13}(u;\lambda) \wt\L_{31}(u;\lambda).
\end{gather*}
\end{remark}

\subsection[The $\slt$ elliptic Gaudin model]{The $\boldsymbol{\slt}$ elliptic Gaudin model}
\label{sec24}

Let us relate the generating function $Q(u,\partial_u)$ in $n=2$ case with the generating function describing the $\mathfrak{sl}_2$ elliptic Gaudin model~\cite{EFR, FS, GZZ}. The latter one corresponds to the $\mathfrak{gl}_2$ elliptic Gaudin model def\/ined by a representation $\rho\colon U(\mathfrak{gl}_2[z,z^{-1}])\to\mathfrak R_0$ such that $\rho\colon e^+_{11}(u)+e^+_{22}(u)\mapsto0$. For example, $\rho=(p\otimes\cdots\otimes p)\circ\rho_{\{v\}}$, where $p(e_{11}+e_{22})=0$, $p(e_{11}-e_{22})=e_{11}-e_{22}$, $p(e_{12})=e_{12}$, $p(e_{21})=e_{21}$. Then the $L$-operator takes the form
\begin{gather*}
\wt\L(u;\lambda)=\left(
\begin{array}{cc}
h^+(u)/2 & f_\lambda^+(u)\\
e_\lambda^+(u) & -h^+(u)/2
\end{array}
\right),
\end{gather*}
where $\lambda=\lambda_{12}=\lambda_1-\lambda_2$ and the half-currents are
\begin{gather*}
h^+(u) =e^+_{11}(u)-e^+_{22}(u)=\sum_{s=1}^N\frac {\theta'(u-v_s)}{\theta(u-v_s)} \big(e_{11}^{(s)}-e_{22}^{(s)}\big), 
\\
e^+_{\lambda}(u) =e^+_{12}(u;\lambda)=\sum_{s=1}^N\frac {\theta(u-v_s+\lambda)}{\theta(u-v_s)\theta(\lambda)} e_{12}^{(s)}, \\
f^+_{\lambda}(u) =e^+_{21}(u;\lambda)=\sum_{s=1}^N\frac {\theta(u-v_s-\lambda)}{\theta(u-v_s)\theta(-\lambda)} e_{21}^{(s)}.
\end{gather*}
Since the $L$-operator depends only on the dif\/ference $\lambda=\lambda_1-\lambda_2$ we can restrict $Q(u,\partial_u)$ to the space $\mathbb A=\{a\in\mathbb{A}_2\mid(\partial_{\lambda_1}+\partial_{\lambda_1})a=0\}\subset\mathbb{A}_2$ coinciding with $\mathbb C((\lambda_{12}))$. Let $\A=\mathfrak R_0\otimes\mathbb A[\partial_{\lambda}]$ then the values of the functions $s_m(u)$ belong to ${\cal N}={\mathfrak N}_{\A}(\mathfrak h)=\{x\in\A\mid hx\in\A h\}$. Since $\rho\colon h_1+h_2\to0$ the operator $\wh D_\lambda$ in this representation has the form $H\partial_\lambda$, where $H=E_{11}-E_{22}$.

Now we calculate the column determinant~\eqref{Q_Ltilde}:
\begin{gather*}
 Q(u,\partial_u)=\det\left(\frac{\partial}{\partial u}-\wh D_\lambda+\wt\L(u;\lambda)-\frac{\theta'(\lambda)}{\theta(\lambda)}\frac{h}{2}\right) \\
\phantom{Q(u,\partial_u)}{}
=\det
\begin{pmatrix}
 \frac{\partial}{\partial u} -\partial_{\lambda}+h^+(u)/2-\frac{\theta'(\lambda)}{\theta(\lambda)}h/2 & f_\lambda^+(u)\\
 e_\lambda^+(u) & \frac{\partial}{\partial u} + \partial_{\lambda}-h^+(u)/2-\frac{\theta'(\lambda)}{\theta(\lambda)}h/2
\end{pmatrix}  \\
\phantom{Q(u,\partial_u)}{}
=\left(\frac{\partial}{\partial u}\right)^2-\frac{\theta'(\lambda)}{\theta(\lambda)}h\frac{\partial}{\partial u}-S_\lambda(u),
\end{gather*}
where $h=h_1-h_2$ and $S_\lambda(u)$ is the following ${\cal N}$-valued function
\begin{gather*} 
 S_\lambda(u)=\big(\partial_{\lambda}-h^+(u)/2\big)^2+\partial_u h^+(u)/2+e_\lambda^+(u)f_\lambda^+(u)
 \qquad {\rm mod} \ \ \A h,
\end{gather*}
which commutes with itself: $[S_\lambda(u),S_\lambda(v)]=0$ mod $\A h$. Using the commutation relation $[e^+_\lambda(u),f^+_\lambda(u)]\hm=-\frac{\partial}{\partial u}h^+(u)+\big(\frac{\theta'(\lambda)}{\theta(\lambda)}\big)'h$ one obtains the generating function for the $\slt$ elliptic Gaudin Hamiltonians:
\begin{gather*}
S_\lambda(u)=\big(\partial_{\lambda}-h^+(u)/2\big)^2
+\big(e_\lambda^+(u)f_\lambda^+(u)+f_\lambda^+(u)e_\lambda^+(u)\big)/2 \qquad {\rm mod} \ \ \A h.
\end{gather*}
This generating function was considered and used in~\cite{EFR,FS} to f\/ind the eigenfunctions of this model.

\subsection{Quantum powers for the Gaudin model}
\label{sec25}

The quantum powers for the elliptic Gaudin models are def\/ined in terms of the operator $\L_D(u)=\L(u;\lambda)-\wh D_\lambda$.
The f\/irst two quantum powers of this operator coincide with its ordinary powers: $\L_D^{\bf[0]}(u)=1$, $\L_D^{\bf[1]}(u)=\L_D(u)$. The others are def\/ined by the following recursive formula
\begin{gather*}
 \L_D^{\bf[k+1]}(u)=\L_D(u)\L_D^{\bf[k]}(u)+\frac{\partial\L_D^{\bf[k]}(u)}{\partial u}.
\end{gather*}
Considering the Newton identities for the Manin matrix ${\cal M}=\frac{\partial}{\partial u}+\L_D(u)=\frac{\partial}{\partial u}+\L(u;\lambda)-\wh D_\lambda$ we conclude that the traces of the quantum powers form an alternative commuting family of functions for the elliptic Gaudin model:
\begin{gather*}
 \big[\tr\big(\L_D^{\bf[k]}(u)\big),\tr\big(\L_D^{\bf[m]}(v)\big)\big]=0 \qquad {\rm mod} \ \ \A_n\mathfrak h.
\end{gather*}
The proof of this formula is the same as for the rational Gaudin model. See details in~\cite{ChF}.

Instead of the quantum powers of $\L_D(u)$ one can consider ``simplif\/ied'' quantum powers $\L_D(u)\L_D^{\bf[m-1]}(u)$. Their traces also form the commuting family of functions def\/ining the same integrable system:
\begin{gather*}
 \big[\tr\big(\L_D(u)\L_D^{\bf[k-1]}(u)\big),\tr\big(\L_D(v)\L_D^{\bf[m-1]}(v)\big)\big]=0 \qquad {\rm mod} \ \ \A_n\mathfrak h.
\end{gather*}
In particular, this family includes the trace of the ordinary second power
\begin{gather*}
 \tr\big(\L_D(u)^2\big)=\tr\big((\L(u;\lambda)-\wh D_\lambda)^2\big).
\end{gather*}

\section{Conclusion}

In Section~\ref{QA} we have constructed a commutative family starting with the Felder $R$-matrix. All results of that section are valid if we would start with another dynamical $R$-matrix satisfying $R(-\hbar;\lambda)=B(\lambda)A^{(12)}$ for some invertible matrix $B(\lambda)$. The universality of this method also implies that one can consider an arbitrary $L$-operator satisfying dynamical $RLL$-relations. In other words, it can be applied to a wide class of models which are described by using a dynamical $R$-matrix. Moreover, the integrals of motion for the models corresponding to a non-dynamical $R$-matrix can be obtained in the same way, or more precisely as it was done in~\cite{T} for the rational case. In particular, it should work for the $XYZ$-model or other models described by the Belavin $R$-matrix.

The use of the Manin matrices is important to provide the universality of our method due to the universality of that notion. First, the characteristic polynomials of Manin matrices are key objects to obtain the commutative family of the (elliptic) $\gln$ Gaudin model by the degeneration. Then the Manin matrices are used to relate the constructed commutative families with another important class of commutative families -- the traces of the quantum powers. The Manin matrices and their characteristic polynomials f\/ind their applications to many important problems in the theory of integrable systems. It is also convenient to use the properties of Manin matrices to prove commutativity of the constructed families.

Analysing the trigonometric limit we have noticed that $L$-operators related with the Hopf algebra $U_q(\wh{\gln})$, such as the $L$-operator of the $XXZ$-model, gives also a Manin matrix after some simple twisting. This means that the corresponding quantum determinants, characteristic polynomials, quantum powers and Newton identities can be written in terms of the ordinary anti-symmetrizers $A^{(1,\ldots,m)}$ by using the theory of Manin matrices, while consideration of the non-twisted $L$-operators leads to the using of $q$-deformations of these anti-symmetrizers and of a $q$-analogue of the Manin matrices.

In  Section~\ref{GAUD} we showed that the degeneration of each dynamical elliptic operator $L(u;\lambda)$ gives a commutative family for the corresponding degenerated $L$-operator $\L(u;\lambda)$. To justify the obtained results for a given $\L(u;\lambda)$ one have to prove the existence of the corresponding $L(u;\lambda)$. We considered the $L$-operators $\L(u;\lambda)$ def\/ined by some homomorphism $\rho\colon U(\gln[z,z^{-1}])\to\mathfrak R_0$ or a representation $\rho\colon U(\gln[z,z^{-1}])\to\End V$. We discuss the existence of the corresponding $L$-operators $L(u;\lambda)$ presenting them explicitly or referring to other works.

 For the representations $\rho=\Pi_{\{v\}}$ we present the corresponding $L(u;\lambda)$ using the notion of the small elliptic quantum group suggested in~\cite{TV}. To obtain the commutative family in the case of arbitrary homomorphism $\rho\colon U(\gln[z,z^{-1}])\to\mathfrak R_0$ it is suf\/f\/icient to construct $L(u;\lambda)$ corresponding to $\rho=\id$. In the $\mathfrak{gl}_2$ (technically $\slt$) case the $L$-operator $L(u;\lambda)$ corresponding to $\rho=\id$ in fact constructed in~\cite{EF}. For the general $\gln$ (technically $\mathfrak{sl}_n$) case the (non-twisted) $L$-operator $L(u;\lambda)$ corresponding to $\rho=\id$ is written in~\cite{KK} as a conjecture.
One also need to relate the $R$-matrix using in~\cite{KK} with the Felder $R$-matrix by some twist as in Subsection~\ref{sec13}. The existence this twist can be quite easily established using the ansatz ${\cal F}=\exp\Big(\frac\hbar2\sum_{i\ne j}\varphi_{ij}(\lambda)\hat h_i\otimes\hat h_j\Big)$, where $\varphi_{ij}(\lambda)=-\varphi_{ji}(\lambda)$ are $\mathbb C[[\hbar]]$-valued functions. Let us f\/inally remark that the degeneration of this twist gives the classical twist considered in Subsection~\ref{sec23}: ${\cal F}=1+\frac\hbar2\mathfrak f+o(\hbar)$.

\subsection*{Acknowledgements}

 The work of V.R.\ was partially supported by the RFBR grant 08-01-00667 and grant of Support for the Scientif\/ic Schools 3036.2008.2. V.R.\ is also grateful to French-Ukrainian PICS 2009. D.T.\ and A.S.\ would like to thank LAREMA and Mathematical Department of the University of Angers for the hospitality during the visits where the part of the work was done.
The work of A.S.\ was partially supported by the EPSRC grant EP/F032889/1 and the RFBR grant 09-01-00239-a.
This work of D.T.\ was partially supported by the Federal Nuclear Energy Agency of Russia, the RFBR grant 07-02-00645, the grant of Support for the Scientif\/ic Schools 3035.2008.2, and the fund ``Dynasty''. V.R.\ and D.T.\ are thankful to the MATPYL
Federation support of D.T.\ visit in Angers. V.R.\ and D.T.\ acknowledge the hospitality and an excellent working atmosphere during the International Workshop ``Elliptic integrable systems, isomonodromy problems, and hypergeometric functions'' in the Hausdorf\/f Centrum and Max-Planck-Institute for Mathe\-matics in Bonn in July 2008. They are grateful to organizers for the invitation. A part of results was presented by V.R.\ in his talk during the MISGAM-SISSA International Conference ``From integrable structures to topological strings and back'' in September 2008. He is thankful to MISGAM and SISSA for this invitation and support. At last but not the least our thanks to anonymous referees for their careful reading and numerous useful remarks and suggestions.

\pdfbookmark[1]{References}{ref}
\LastPageEnding

\end{document}